\documentclass[conference, 10pt]{IEEEtran}
\IEEEoverridecommandlockouts
\usepackage{cite}
\usepackage{amsmath,amssymb,amsfonts}
\usepackage{algorithmic}
\usepackage{graphicx}
\usepackage{textcomp}
\usepackage{xcolor}
\usepackage{comment}
\def\BibTeX{{\rm B\kern-.05em{\sc i\kern-.025em b}\kern-.08em
    T\kern-.1667em\lower.7ex\hbox{E}\kern-.125emX}}
\begin{document}

\title{Resilience of IOTA Consensus\\
}

\author{ \IEEEauthorblockN{Hamed Nazim Mamache}
\IEEEauthorblockA{
\textit{Sorbonne University},
Paris, France \\
hamed\_nazim.mamache@etu.sorbonne-universite.fr}
\and

\IEEEauthorblockN{Gabin Mazué}
\IEEEauthorblockA{
\textit{Sorbonne University},
Paris, France \\
gabin.mazue@etu.sorbonne-universite.fr}
\and

\IEEEauthorblockN{Osama Rashid}
\IEEEauthorblockA{
\textit{Sorbonne University},
Paris, France \\
osama.rashid@etu.sorbonne-universite.fr}
\and
\IEEEauthorblockN{Gewu Bu}
\IEEEauthorblockA{ 
\textit{University Clermont Auvergne}\\
Clermont Auvergne, France \\
gewu.bu@uca.fr} 
\and
\IEEEauthorblockN{Maria Potop-Butucaru}
\IEEEauthorblockA{
\textit{Sorbonne University},
Paris, France \\
maria.potop-butucaru@lip6.fr}
}

\maketitle

\begin{abstract}
Blockchains are appealing technologies with various applications ranging from banking to networking. IOTA blockchain is one of the most prominent blockchain specifically designed  for IoT environments.  {\color{red}In this paper we investigate the convergence of two Consensus proposed by IOTA: Fast Probabilistic Consensus and Cellular Consensus,} when run on top of various topologies. Furthermore, we investigate their resilience to various types of adversaries. Our extensive simulations confirm that both Fast Probabilistic Consensus and Cellular Consensus have poor convergence rates even under low power adversaries and have poor scaling performances except for the case of Watts Strogatz topologies. Our study points out that the design of IOTs dedicated blockchains is still an open research problem and gives hints design. {\color{red}This why the foundation IOTA works on a complete version of consensus, Coordicide, for the new IOTA, while regarding these two as components of.}     
 
\end{abstract}


\section{Introduction}
\label{sec:typesetting-summary}
Internet of Things (IoT) devices are used in a large range of applications such as smart grids, smart foraging, smart buildings, smart supply chains or smart medical applications
and the IoT environment \cite{IGMVQIFT16} is expected to further expand  even more ubiquitous deployment thanks to the fifth generation of networks (5G) \cite{ABCHLSZ14} and beyond.  

Due to the vulnerabilities of IoT to various attacks and the very harmful potential consequences, the currently dominating approach in the management of IoT devices is centralizing control operations at IoT gateways, which are considered as the natural function to absolve access control, data filtering and mixing operations. However, centralization does not appear viable when one envisions hundreds of thousands of devices per km2 or per cell \cite{IGMVQIFT16}, especially when those devices can be constrained in size and power supply. 

The use of Distributed Ledger Technologies (DLT) can respond to both security and decentralization needs in the management of IoT devices. Distributed Ledger Technologies (DLT) such as blockchains provide a secure way to share information between a high number of independent nodes operating under different authorities, while ensuring high availability and immutability.
Distributed Ledger Technology pioneered by Bitcoin  technology created a new design philosophy for executing and storing transactions in a decentralized and secure fashion \cite{nakamoto2008bitcoin}. 

 A blockchain is a distributed ledger that mimics the functioning of a classical traditional ledger (i.e. transparency and falsification-proof of documentation) in an untrusted environment where the computation is distributed.   Traditional blockchain systems such as Bitcoin \cite{nakamoto2008bitcoin} or Ethereum \cite{Ethereum} maintain a continuously-growing list of ordered blocks that include one or more transactions that have been verified by the members of the system, called miners. Blocks are linked using cryptography and the order of blocks in the blockchain is the result of a form of agreement (consensus) among the system participants.  Bitcoin technology and similar proposals (e.g Ethereum) came with several drawbacks that prevent them from being used as standards for IoT industry.    Therefore, alternative solutions have been opened by  IOTA  \cite{popov2016tangle}. IOTA's data structure is a \emph{Directed Acyclic Graph} (DAG) based distributed ledger, also known as the \emph{Tangle}, aimed to overcome limitations of Bitcoin-like distributed ledgers when used in IoT environment while preserving equivalent security levels. Transactions are continuously appended to the tangle. Similar approaches have been proposed by Spectre or Phantom \cite{DBLP:journals/iacr/SompolinskyLZ16,DBLP:journals/iacr/SompolinskyZ18}.
However,  IOTA and similar approaches have not yet been adopted by the IoT industry because of 1) lack of strong consistency guarantees and 2) unclear resistance to attacks.   In order to respond to these criticism IOTA proposed recently in \cite{coordicide} attacks resilient consensus mechanisms  that plugged into the IOTA Tangle will offer strong consistency guarantees.  Two consensus algorithms are proposed: Fast Probabilistic Consensus (FPC) and Cellular Consensus (CC). 
These two proposals have been partially evaluated in \cite{popov2021fpc,capossele2019robustness}. 

In this work we investigate the performances of IOTA consensus in several aspects. First, we run  Fast Probabilistic Consensus (FPC) and Cellular Consensus (CC) on top of various topologies{, \color{red}from theoretical to practical} (2D Grid, Torus and Watts-Strogatz model \cite{watts1998collective}) then we evaluate their  resilience to adversarial behavior. Our evaluation is conducted with OMNET++ simulator enriched with  three adversarial models introduced in \cite{coordicide}.
Even though most of the results reported in our study are negative they contain hits in order to design an efficient IoT dedicated blockchain.

\section{IOTA Distributed Consensus}

In this section, we briefly describe the operating principles of the two   consensus mechanisms proposed by IOTA \cite{coordicide}. {\color{red}The basic idea is all honest nodes in the system should agree dynamically with a commun opinion by a distributed way, so that this opinion cannot be changed by others easily}.

Consider a connected network composed of $N$ nodes, enumerated as $\{1…i…N\}$. Some lossless links connect nodes in the network. Nodes connected directly are \emph{neighbors}. 

We follow the setting proposed in \cite{popov2021fpc,capossele2019robustness}, assuming that the time is discrete and divided into \emph{rounds}.

Each node has an \emph{opinion} status, $O_i(r) \in \{0,1\}$ at the round $r$.  \emph{Consensus} is achieved, if $\forall i, j \in N, O_i(r_{end}) = O_j(r_{end})$, where $r_{end}$ is the last round of simulation.

An opinion held by most of the nodes is a \emph{major opinion}. 
The \emph{convergence rate} is the percentage of runs leading to a  consensus stage given $P0$, where $P0$ is the probability that a node has opinion $0$ at round $0$. If a node does not have opinion $0$ at round $0$ it will have opinion $1$.

\subsection{Fast Probabilistic Consensus (FPC)}
{\color{red}The idea of Fast Probabilistic Consensus algorithm is based on the query/reply of opinion from nodes in the network.} Algorithm executed by each node is as follows:
\begin{itemize}
    {\color{red}\item Query randomly a number of nodes in network at each discrete time round $t$;}
    \item Wait for the chosen nodes to respond and give their opinions;
    \item Calculate the mean of the received opinions.
\end{itemize}
Once the node executing Fast Probabilistic Consensus algorithm has calculated the mean, if it is the first round, he will compare it to a threshold $\tau$, if the mean is bigger than $\tau$, then its opinion becomes 1, otherwise becomes 0. If it is not the first round, the node will generate a random variable $U_t$ following a uniform law in function of $\tau$ between $[\beta,1-\beta]$ (where $\beta$ is the uniform low parameter). If the mean is bigger than $U_t$, the opinion becomes $1$, if it is smaller, the opinion becomes $0$ and if they are equal, the opinion stays the same as the previous round.


{\color{red}In \cite{popov2021fpc}, FPC has two special assumptions. The first is that all nodes in the network have a common random values sequences. This random values sequences should theoretically be provided by a trusted third party in other network layer. Through this common sequence, all nodes will "randomly" select the same threshold, $U_t$, at each execution round $t$. The second assumption is that each node has the complete view of all $N$ nodes in network, and can freely query $Q$ nodes among them. Although the experiments in the \cite{capossele2019robustness} show that even if nodes have only 50\% of the network view, the algorithm still works.}

{\color{red}However, in an IOT network environment where millions of devices can join and exit at any time, neither of these assumptions is realistic. In order to adapt FPC to distributed IOT environment, we made the following adjustments: 1) Remove the assumption about the common random values sequence. 2) The node only knows the nodes directly connected to it, the neighbors. In order to represent the process of randomly querying Q nodes of original FPC, we use the Random Walk method. Through random walks, nodes that only know its neighbors can still query other nodes within the range of random walks that are not directly connected.}

{\color{red}When a node need to query nodes in each round, it launches a number of random walks of distance $D$. When a random walk stops at a node, this node will be chosen as a node to be queried.} When a node launches a random walk within distance $D$, it will choose randomly one of its neighbors and send a message containing a list of visited nodes and a positive distance value, $D-1$, as the distance decremented of $1$. At this point, the list of visited nodes will only have the node that launched the random walk. Once one node receives that message, it will verify whether the distance remaining is greater than $0$. If so, it adds itself in the list of visited nodes, decreases the distance value and sends the message to one of its neighbors. This neighbor is chosen uniformly between the neighbors that have not already been visited so that no random walk can go through the same node twice. If all of the neighbors have already been visited, no random walk will be launched. A node stops launching the random walk if the remaining distance value reaches $0$.



\subsection{Cellular Consensus(CC)}
In Cellular Consensus(CC), each node acts as an individual  agent, that changes its opinion in case of conflict with its neighbors and adopts the major opinion among its neighbors. 

At the beginning of each round, every node sends a "heartbeat" of its signed current opinion and the opinions from the previous round of his neighbors, each one signed by the issuing node.

When a node receives an opinion given by one of its neighbors, it will evaluate this opinion by a "proof" accompanying the opinion. This "proof" is materialized by the opinions of the neighbor’s neighbors. That will allow nodes to monitor each other and to detect if someone is lying independently of its neighbors. If this "proof" shows that the neighbor is lying, it will immediately be blacklisted by the node and none of his opinions will be taken into account.

Since the previous opinions of the neighbors cannot be faked, every node can validate that the received opinion is indeed correct.

In detail, the cellular consensus algorithms work as follows.
At each step of the algorithm, each node holds an opinion, which can be $0$, $1$ or a temporary opinion state $-1$.

If the major opinion among neighbors of node $i$ in round $r$ is: 
\begin{itemize} 
\item $0$, then the opinion of node $i$ in round $ r + 1 $ will be $ 0 $,
\item $1$, then the opinion of node $i$ in round $ r + 1 $ will be $ 1 $,
\item non-existent (i.e. there is no majority opinion), then the opinion of node $i$ in round $ r + 1$ will be $ -1 $.
\end{itemize}

\section{Topology Presentation}
In this section we introduce three network typologies that used in our experimentation: {\color{red}{\bf 2D Grid}, {\bf Torus} {\color{red}, two topologie theoretical often used as reference topologie in algorithm analyzing,} and {\bf Watts-Strogatz} \cite{watts1998collective}, a realistic network model which is well-suitable for IoT.}

\subsection{2D Grid}
The first topology studied is the simplest of all. We consider a grid or matrix, which contains nodes. Each nodes' neighbors will be those adjacent to them. Their number therefore varies according to the position of the node in the grid. Indeed, if a node ends up in a corner, it will only have 2 neighbors. Also, if it ends up on an edge, it will have 3 neighbors. Finally, if this one is in the center of the grid, it will have 4 neighbors.

\subsection{Torus}
The second topology studied is an improvement of the 2D grid and avoids getting stuck in the corners and on the edges. To do this we will transform our grid into a torus \cite{Contrepoints}. We therefore connect the top and bottom edges together, and repeat the same process for the right and left edges. By connecting the edges between them, we add neighbors to the nodes which are in the corners and on the edges. Indeed, if we are in the top left corner, we add the bottom left corner and the top right corner to the list of our neighbors. If we are on an edge, we add the node which is on the opposite edge. Finally, if we are located in the center, nothing changes. The number of neighbors is therefore equal to 4 for any node in the network. 


\begin{figure}[htbp]
    \centering
    \includegraphics[scale=0.36]{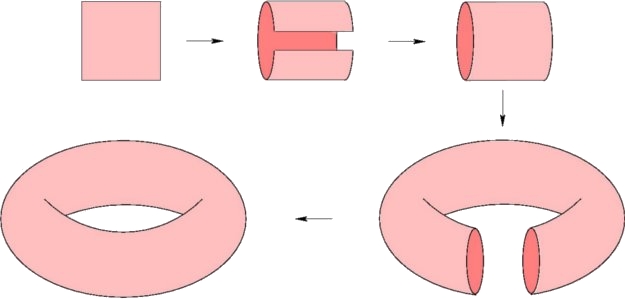}
    \caption{Transformation of a 2D grid into a torus}
    \label{tore}
\end{figure}

\subsection{Watts-Strogatz model graphs}
The Watts-Strogatz model \cite{watts1998collective} is a graph generation model, possessing the small world property (the computation of the shortest path between two nodes is logarithmic). This method takes in parameters $N$ the number of nodes, $K$ the average degree of the nodes in the network and $P$ the probability allowing to change the edges. The goal of the process is to start from a ring graph in order to process every edge of a node. Indeed, each edge can change recipient with the probability P passed as a parameter. 


\begin{figure}[htbp]
    \centering
    \includegraphics[scale=0.9]{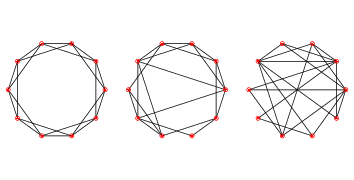}
    \caption{Watts-Strogatz method process}
    \label{watts}
\end{figure}

\section{Byzantines adversaries}
In order to simulate possible attacks and predict critical security cases, three mentioned byzantine adversaries \cite{capossele2019robustness} were implemented:

\subsection{Cautious adversaries.}
These nodes are able to lie on every round of the process with a probability $P_{lying}$. However, the opinion sent during the same round  is always the same even though the queries come from different nodes. 
\subsection{Semi-Cautious adversaries}
These nodes will not lie, however, they may not respond to a query, with a probability $P_{silence}$. Thus delaying the process of convergence and reducing the number of accessible nodes in the network.
\subsection{Berserk adversaries}
This adversary is stronger than the previous two adversaries. It behaves  similar to Cautious, except that it is able to provide different responses to different queries received in the same round. Thus, during the same round, it can send his true opinion, then lie and respond with a wrong opinion. 
 
\section{Simulation results}
In this section we will present a selection of our extensive simulations related to the resilience of IOTA consensus algorithms with various adversaries and different network topologies. 
We use  OMNet++ simulator enriched with  the three topologies and the three adversaries models.  We run our simulations on a physical machine with 8 cores 16 threads and 16 GB RAM. We also run  simulations in a virtual machine with 6 VCPU cores and 16 GB RAM. 


Our extensive simulations show that there is no difference between Berserk adversaries and Cautious adversaries in terms of convergence rate. That is because, in long runs, both malicious nodes will give the same quantity of false information to their neighbors in average. If we consider $M$ the number of rounds and $X$ the number of queries that the malicious node will receive per round on average, a Cautious adversary will lie in $\frac{M}{2}$ rounds which makes an average quantity of lies of $\frac{M \times X}{2}$. As for the Berserk adversary, it will lie in average of $\frac{X}{2}$ regardless of the round which also makes a total of $\frac{MX}{2}$ lies. 
In the following, we will detail only the resilience of Fast Probabilistic Consensus and Cellular Consensus to cautious and semi-cautious adversaries.

\subsection{Fast Probabilistic Consensus resilience}
We list here the basic parameters involved in Fast Probabilistic Consensus.
\begin{itemize}
    \item Distance (lenght) of the random walks: $D=4${\color{red}, a relative small valeur to represent a limited knowledge range of neighbors};
    \item Number of nodes queried $10$; 
    \item Initial threshold : $\tau = 0.5$;
    \item Uniform law parameter : $\beta=0.25$;
    \item $K = 10$, $P = 1$ for Watts-Strogatz model 
    \item $P_{lying} = 50\%$ for Cautious;
    \item $P_{silence} = 50\%$ for Semi-Cautious;
    \item Number of rounds : $M = 30$,
\end{itemize}

\subsubsection{Convergence rate according to the initial division probability $P_0$ for different network sizes without malicious nodes}

\begin{figure}[htbp]
    \centering
    \includegraphics[scale=0.09]{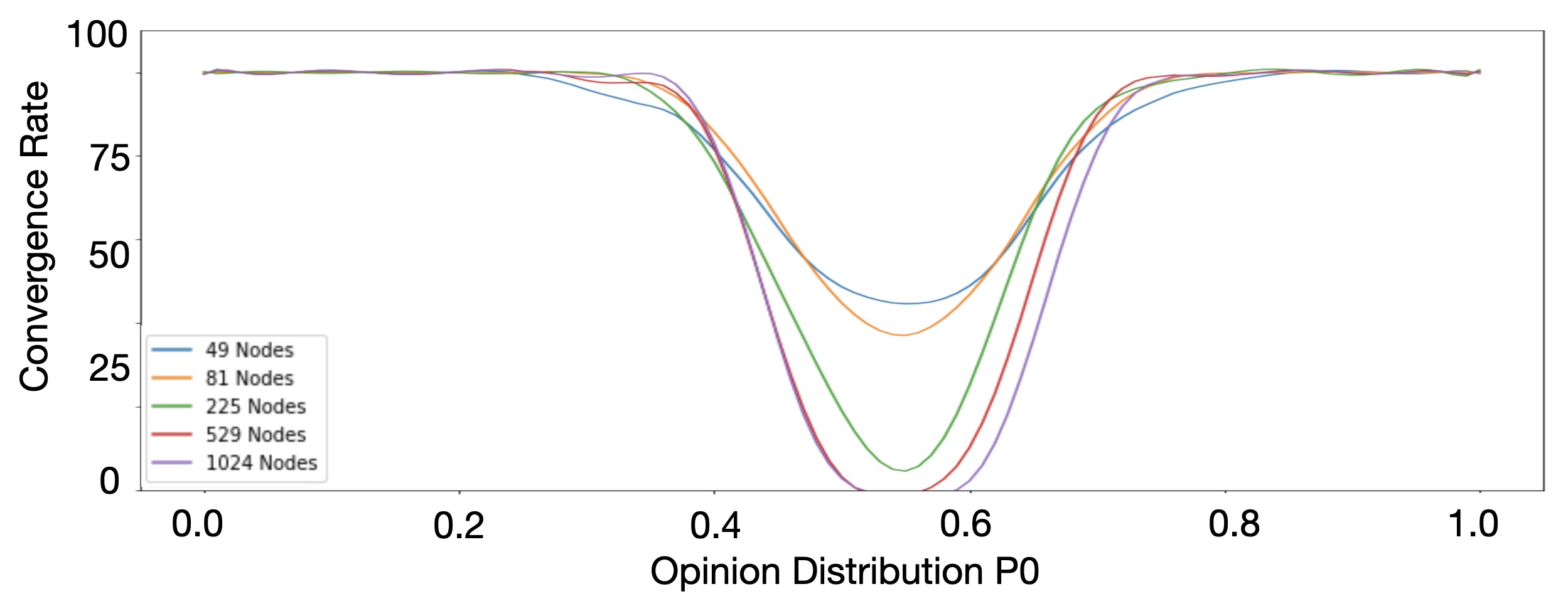} \\
    (a) 2D Grid\\
    \includegraphics[scale=0.09]{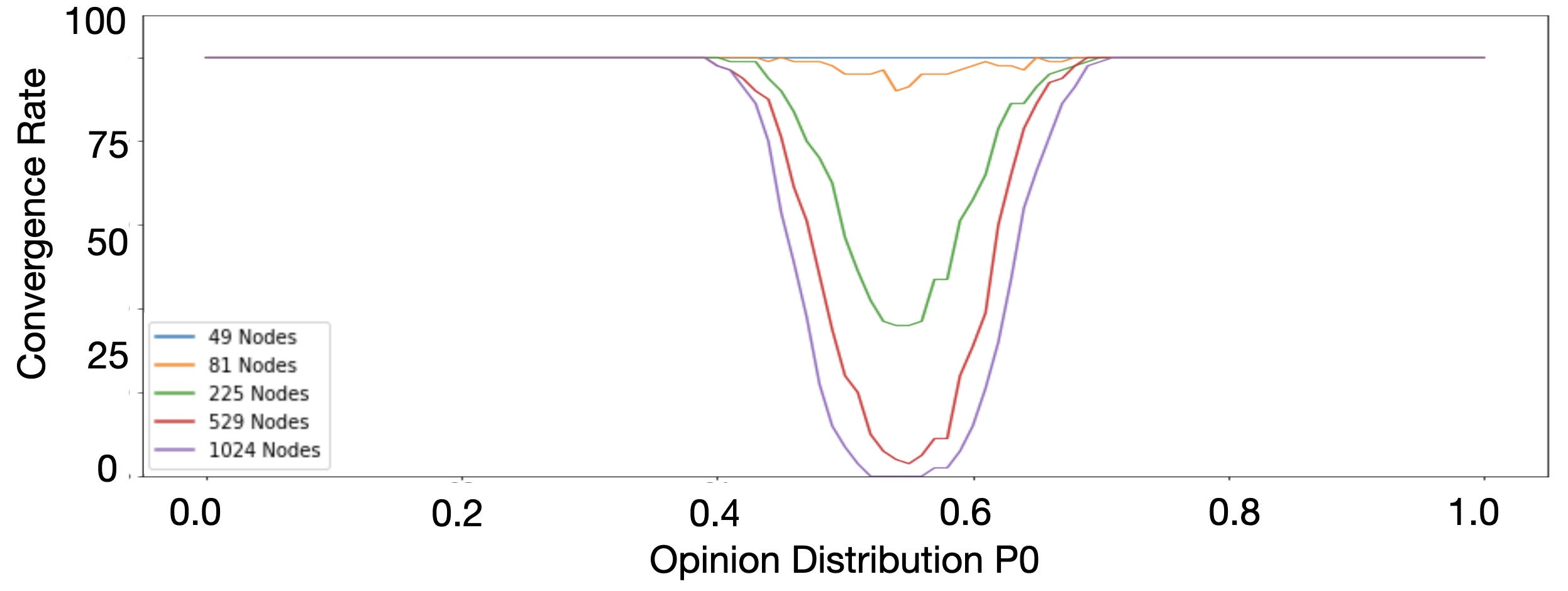}\\
    (b) Torus \\
    \includegraphics[scale=0.09]{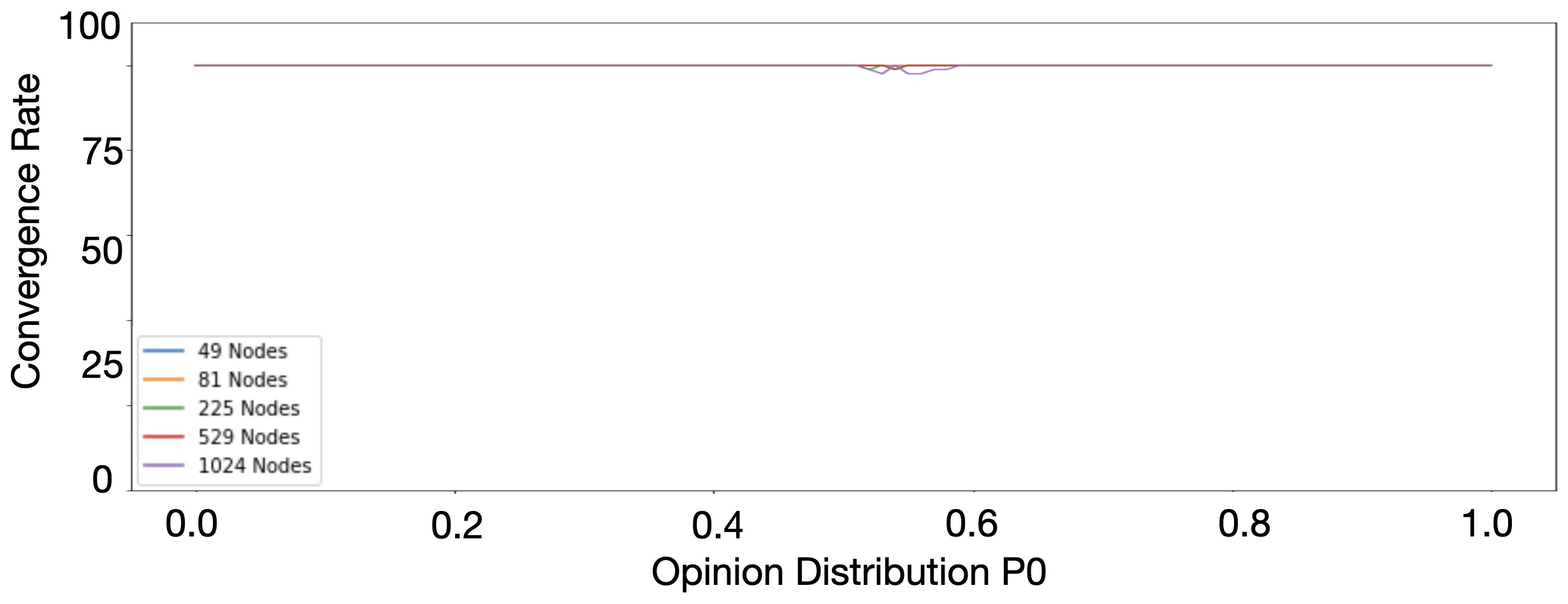} \\
    (c) Watts-Strogatz
    \caption{Convergence rate ccording to the initial division probability $P_0$ for different network sizes, without malicious nodes}
    \label{P1_1}
\end{figure}

We first studied the impact of the number of nodes, $N$, and the initial opinion distribution, $P_0$ on the convergence rate. In Figure \ref{P1_1}, we show the convergence rate function to the initial division probability $P_0$ in each topology and for different network sizes from $N = 49$ to $N = 1024$. In Torus, with small network sizes, the convergence rate is 100\% regardless of the value of $P_0$. While in Grid, even  with small network sizes, the convergence rate drops drastically  when $P_0 \in [0.35, 0.75]$. In terms of network size, for the Grid and the Torus topologies, the more nodes are in network, the lower convergence rate is. Hence, Fast Probabilistic Consensus is not scalable. 
On the contrary, the Watts-Strogatz topology seems to converge well regardless of the number of nodes change. 

\subsubsection{Convergence rate according to the initial division probability $P_0$ for different average number of neighbors $k$ in a Watts-Strogatz graph without malicious nodes}

\begin{figure}[htbp]
    \centering
    \includegraphics[scale=0.09]{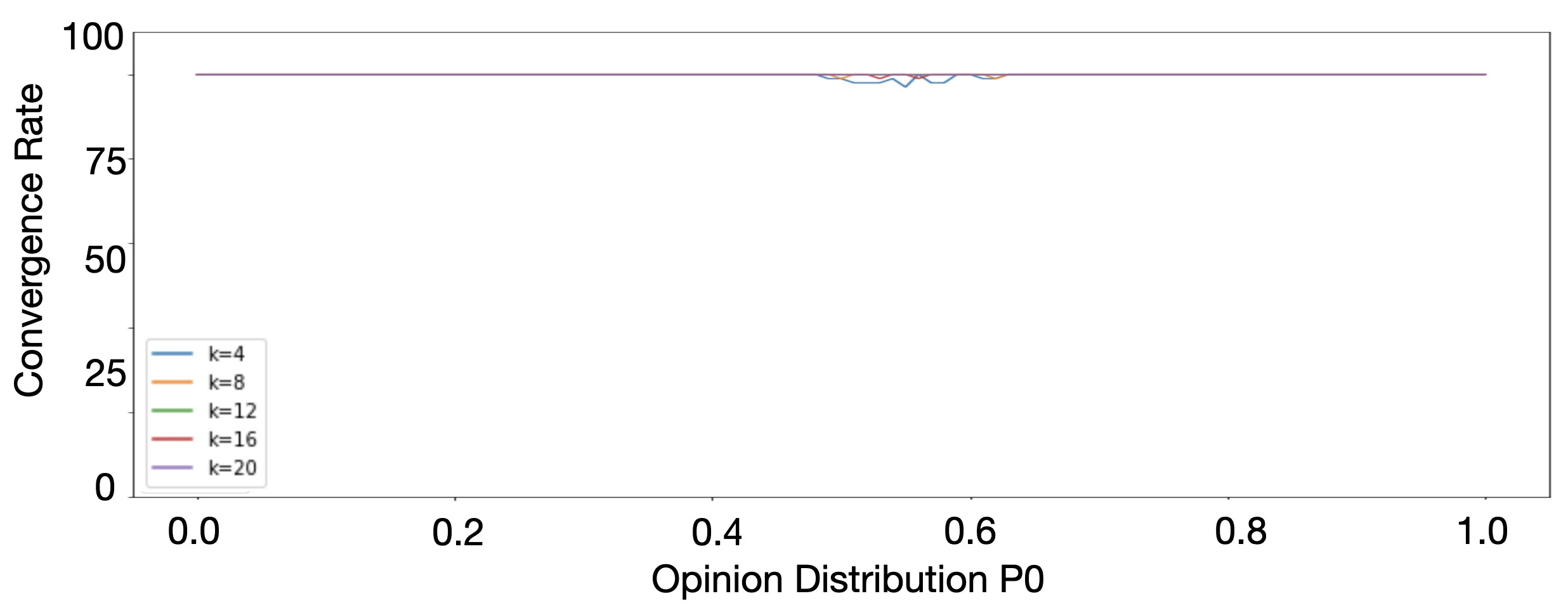}
    \caption{Convergence rate according to the initial division probability $P_0$ for different average number of neighbors $K$ in a Watts-Strogatz without malicious nodes}
    \label{P1_2}
\end{figure}

In Figure \ref{P1_2}, we adjust the average number of neighbors of each node, $K$, from $4$ to $20$. As we can see, if a node has more neighbors, it should converge more easily towards the majority opinion of the network. For $K=20$, the network always converges, however, the network had already a good convergence rate with $K=4$. This implies that the number of direct neighbors of a node in FPC does not mainly affect the convergence rate.

\subsubsection{Convergence rate according to the initial division probability $P_0$ for different number of rounds $M$ without malicious nodes}

\begin{figure}[htbp]
    \centering
    \includegraphics[scale=0.09]{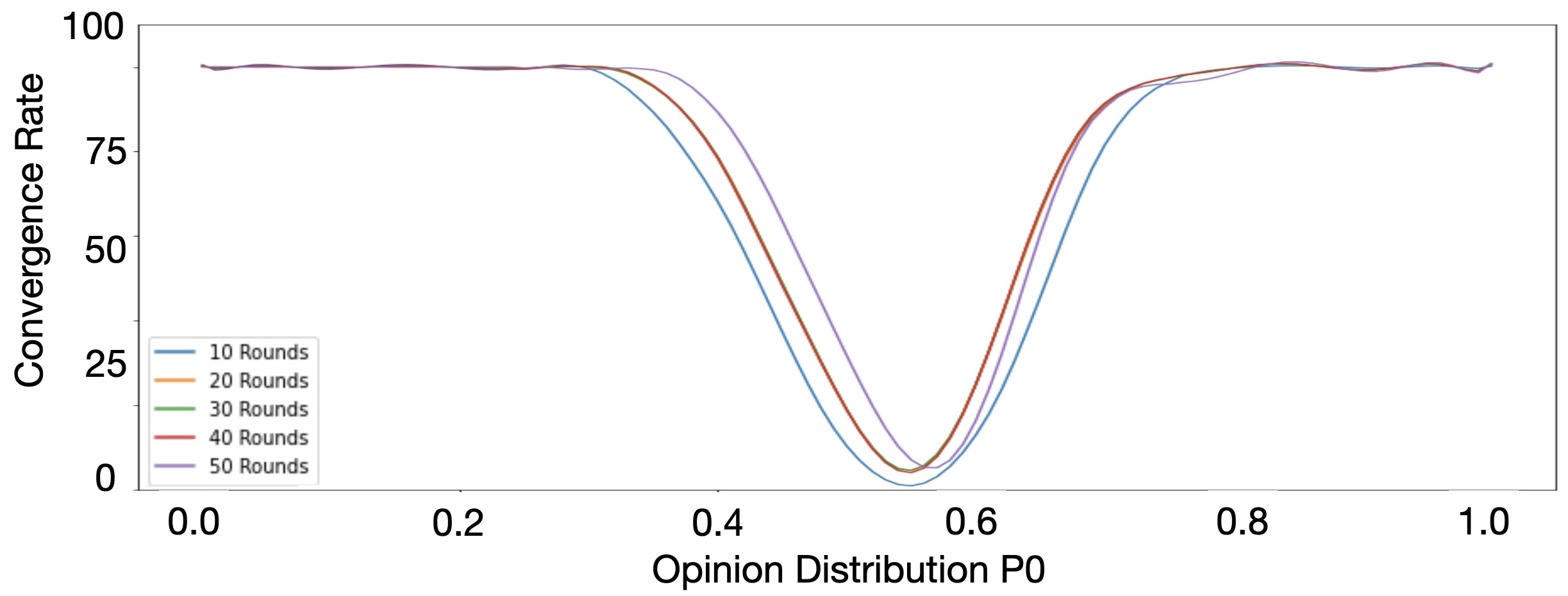}\\
    (a) 2D Grid\\
    \includegraphics[scale=0.09]{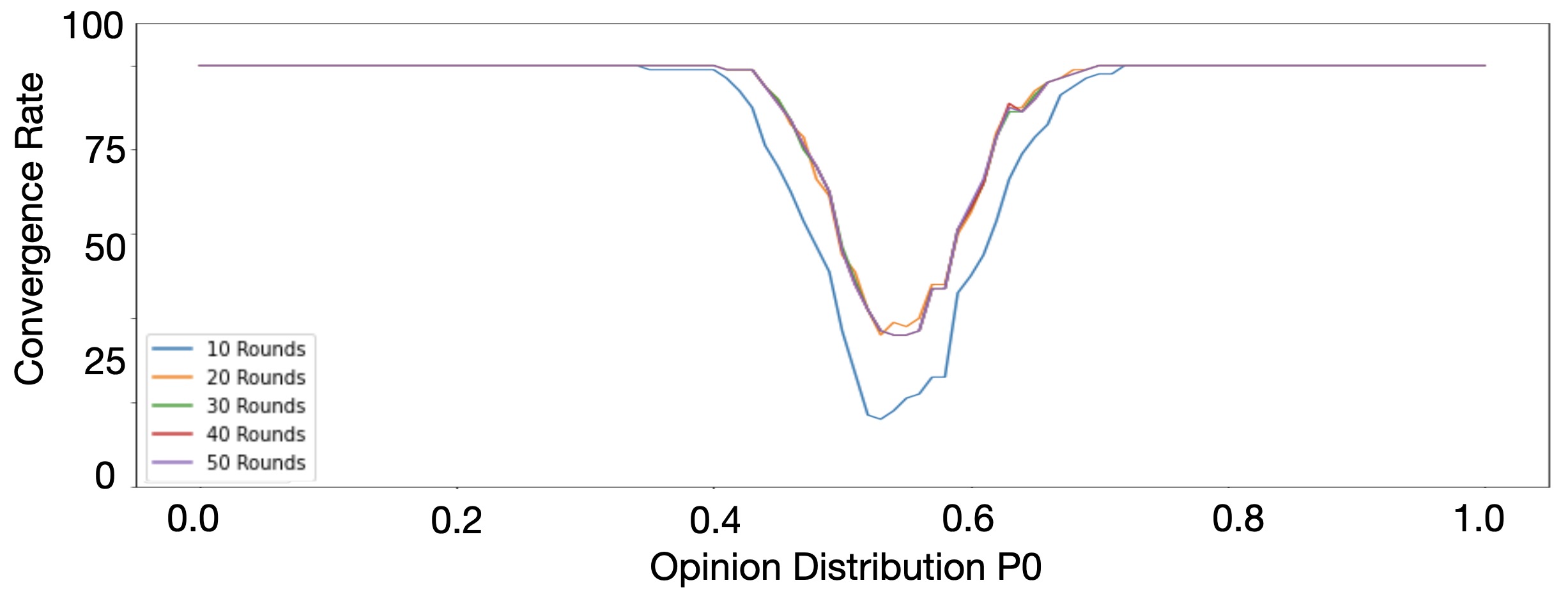} \\
    (b) Torus \\
    \includegraphics[scale=0.09]{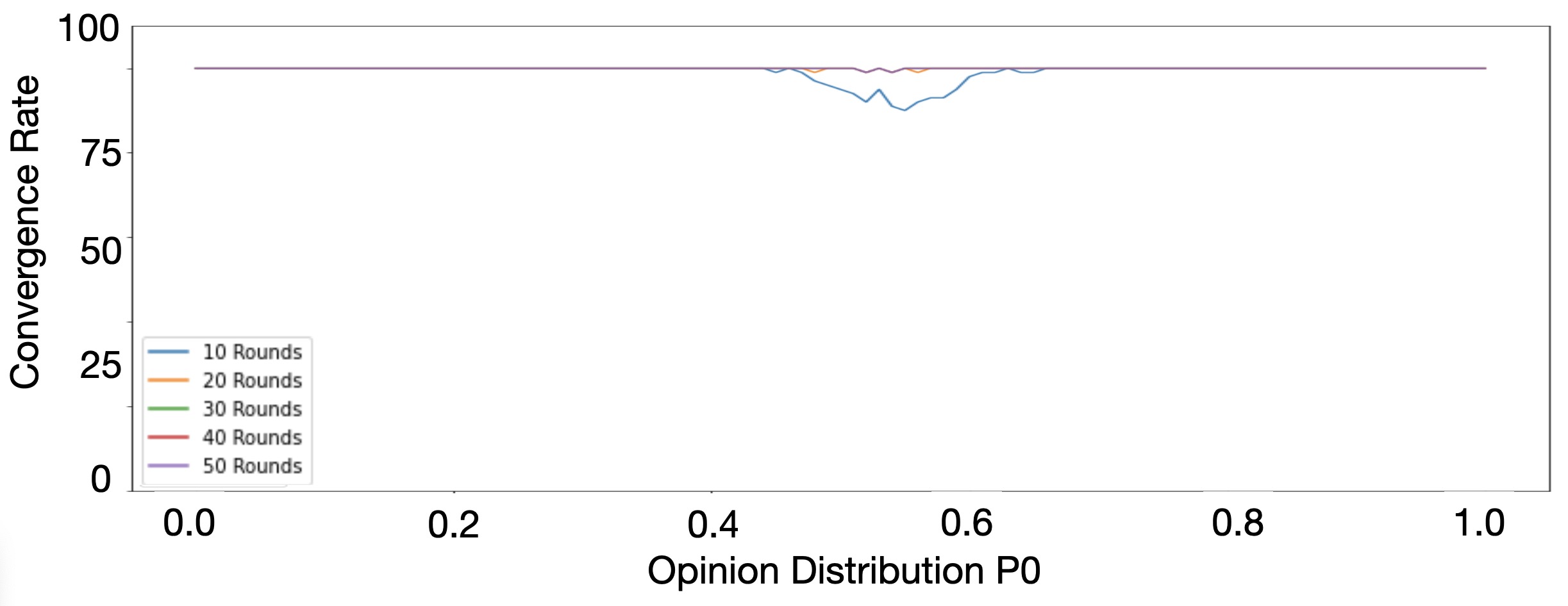} \\
    (c) Watts-Strogatz
    \caption{Convergence rate according to the initial division probability $P_0$ for different number of rounds $M$ without malicious nodes}
    \label{P1_3}
\end{figure}

Figure \ref{P1_3} shows how the number of rounds can affect the convergence rate by setting $M$ from 10 to 50. We can observe that from 20 rounds to 50 rounds, there is no noteworthy difference, that means that for all topologies, 20 rounds is enough to get convergence. In further simulations, we kept 30 rounds just to be sure that the network will have enough time to converge (if it converges). Interestingly, for the rest of simulations, even we introduce adversaries, increasing the execution round will not help at all. 

\subsubsection{Convergence rate according to the initial division probability $P_0$ for different network sizes, with 33\% Cautious Adversaries}

\begin{figure}[htbp]
    \centering
    \includegraphics[scale=0.09]{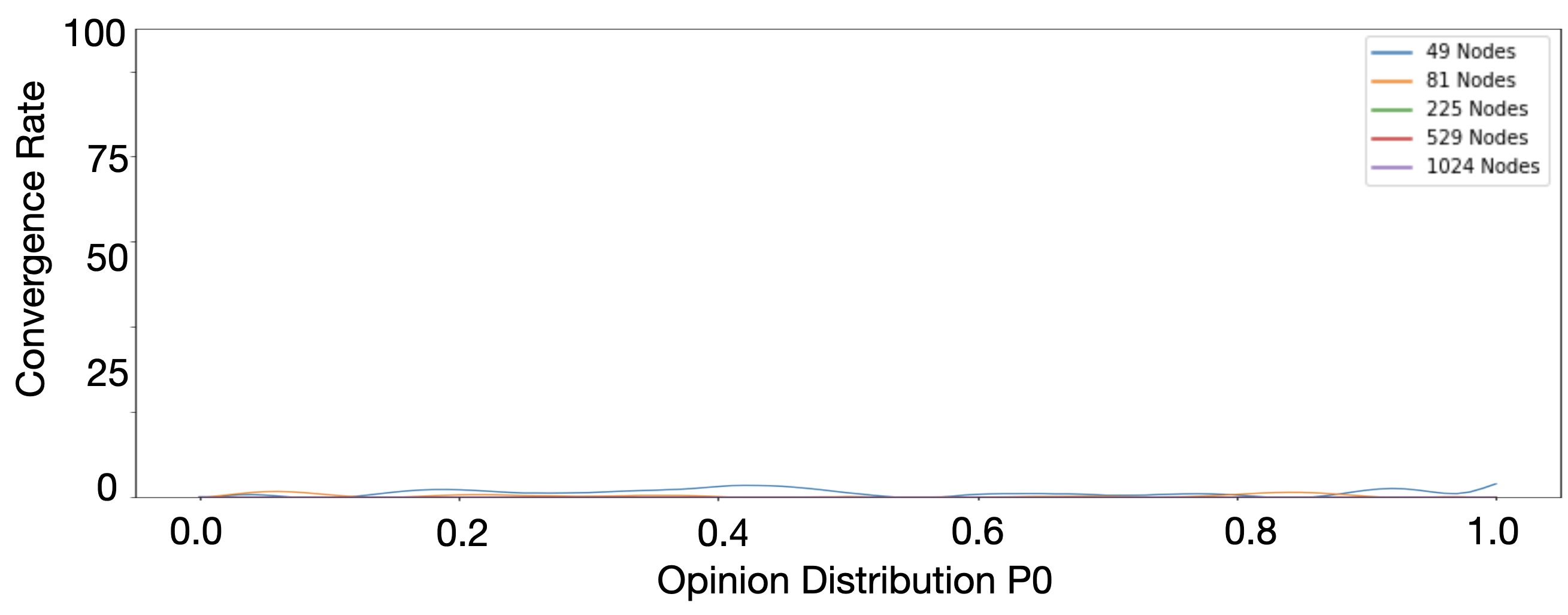}\\
    (a) 2D Grid\\
    \includegraphics[scale=0.09]{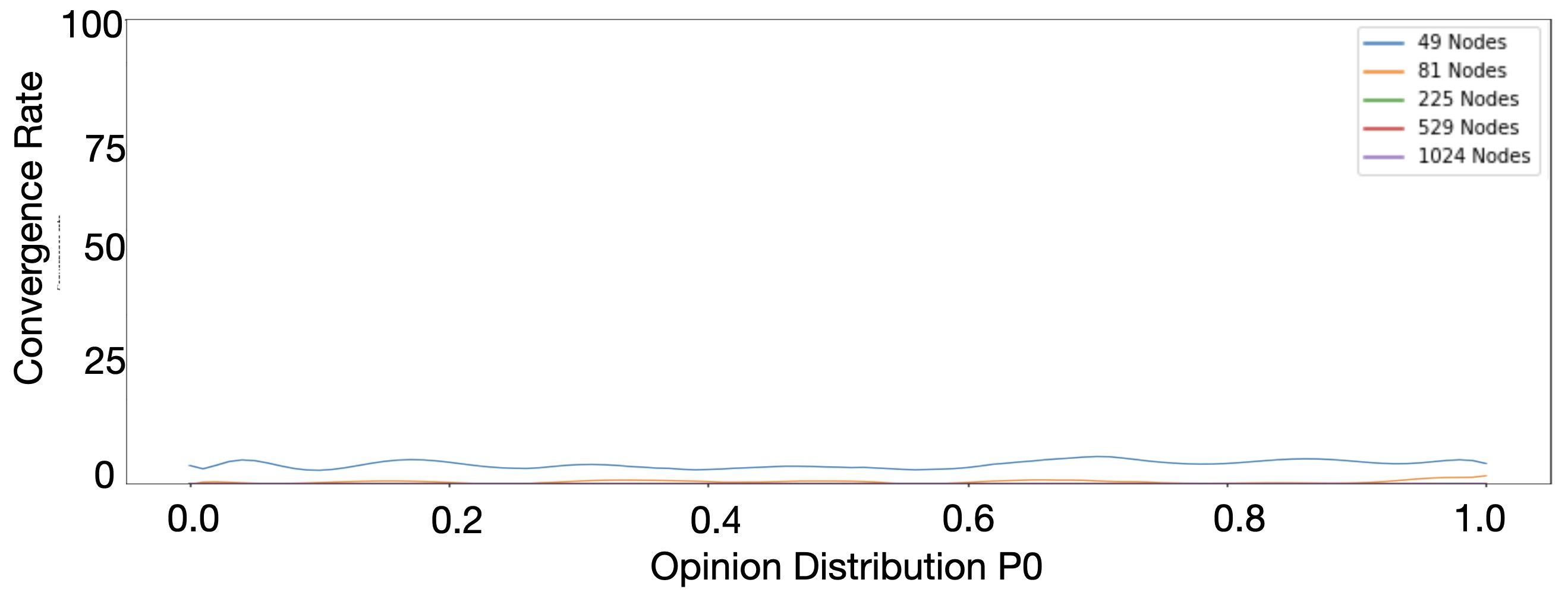} \\
    (b) Torus \\
    \includegraphics[scale=0.09]{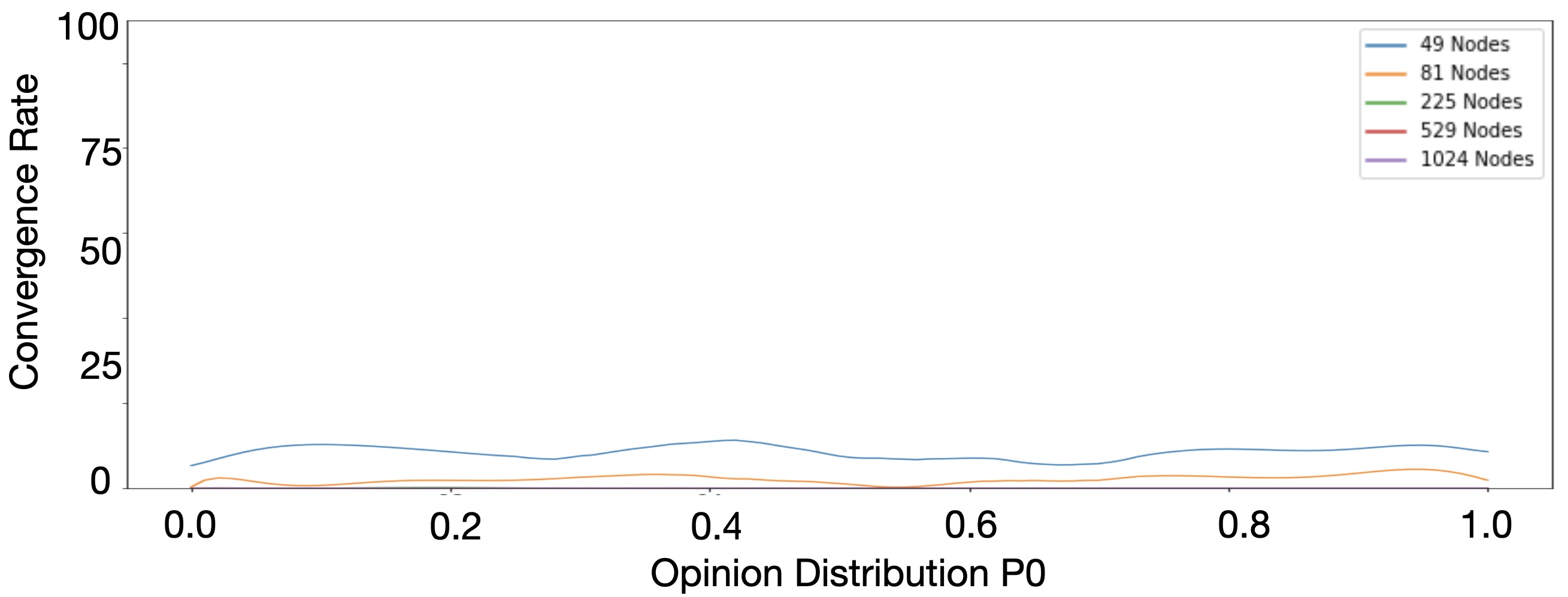} \\
    (c) Watts-Strogatz
    \caption{Convergence rate according to the initial division probability $P_0$ for different network sizes, with 33\% Cautious Adversaries}
    \label{P1_4}
\end{figure}

We inspected first the Fast Probabilistic Consensus resilience when 33\% of total nodes in the network ($N/3$) are Cautious adversaries. This is motivated by the fact that $N/3$ resilience is the upper bound in terms of Byzantine resilience for consensus protocols.

we observed that in Figure \ref{P1_4}, 
the convergence rate drops below $5\%$ for all studied topologies regardless of the initial opinion distribution $P_0$ or the network size $N$. This is due to the fact that the Cautious adversaries can spread freely their lies across the network when nodes ask their opinions via the Fast Probabilistic Consensus queries. Honest nodes therefore cannot make a correct decision. In the following simulations, we fix the network size by choosing a reasonable medium network size, $N = 225$ and study the resilience of Fast Probabilistic Consensus to various percentages of Cautious adversaries.


\subsubsection{Convergence rate according to the initial division probability $P_0$ for different percentages of Cautious adversaries $P_{malicious}$}

\begin{figure}[htbp]
    \centering
    \includegraphics[scale=0.09]{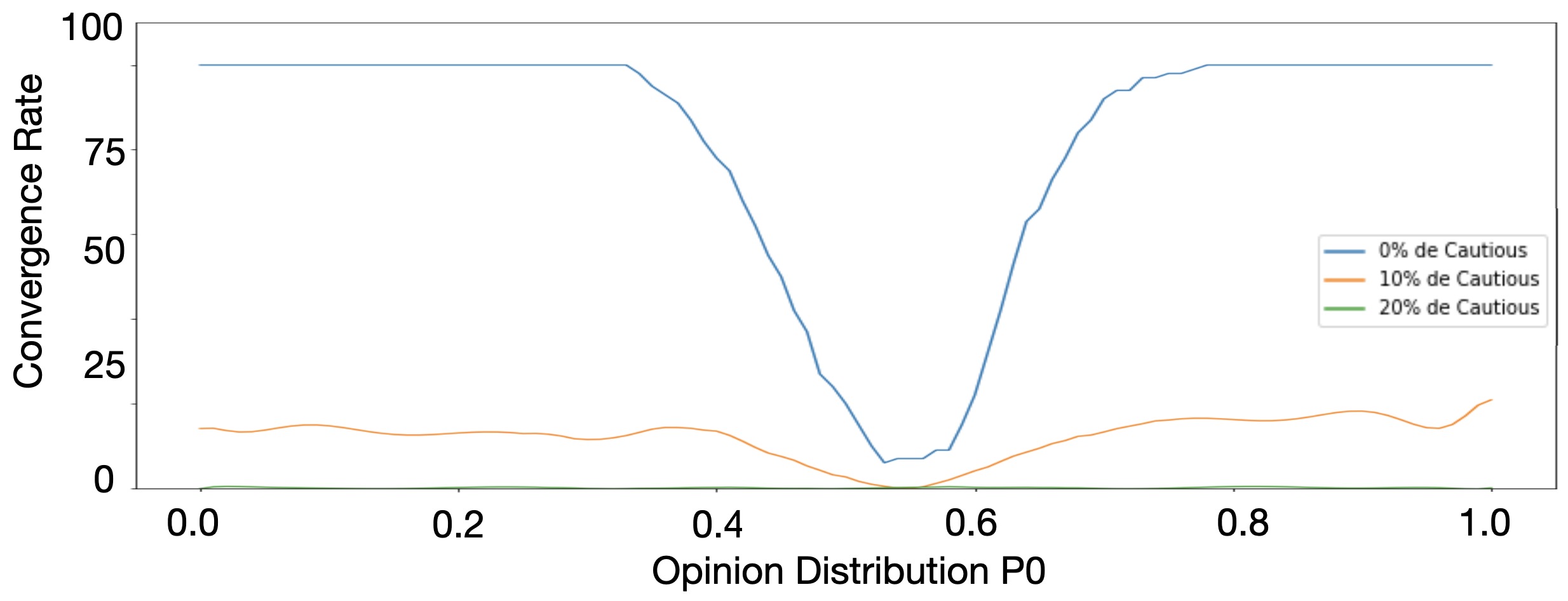}\\
    (a) 2D Grid\\
    \includegraphics[scale=0.09]{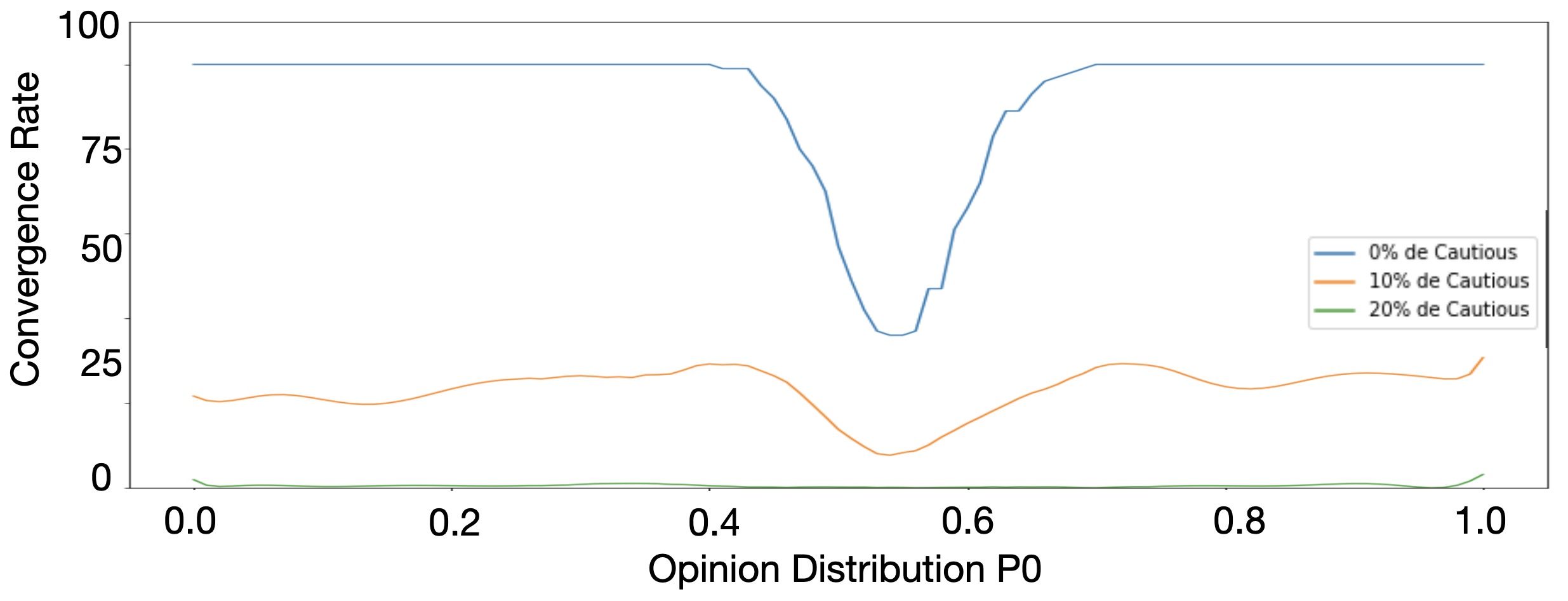} \\
    (b) Torus \\
    \includegraphics[scale=0.09]{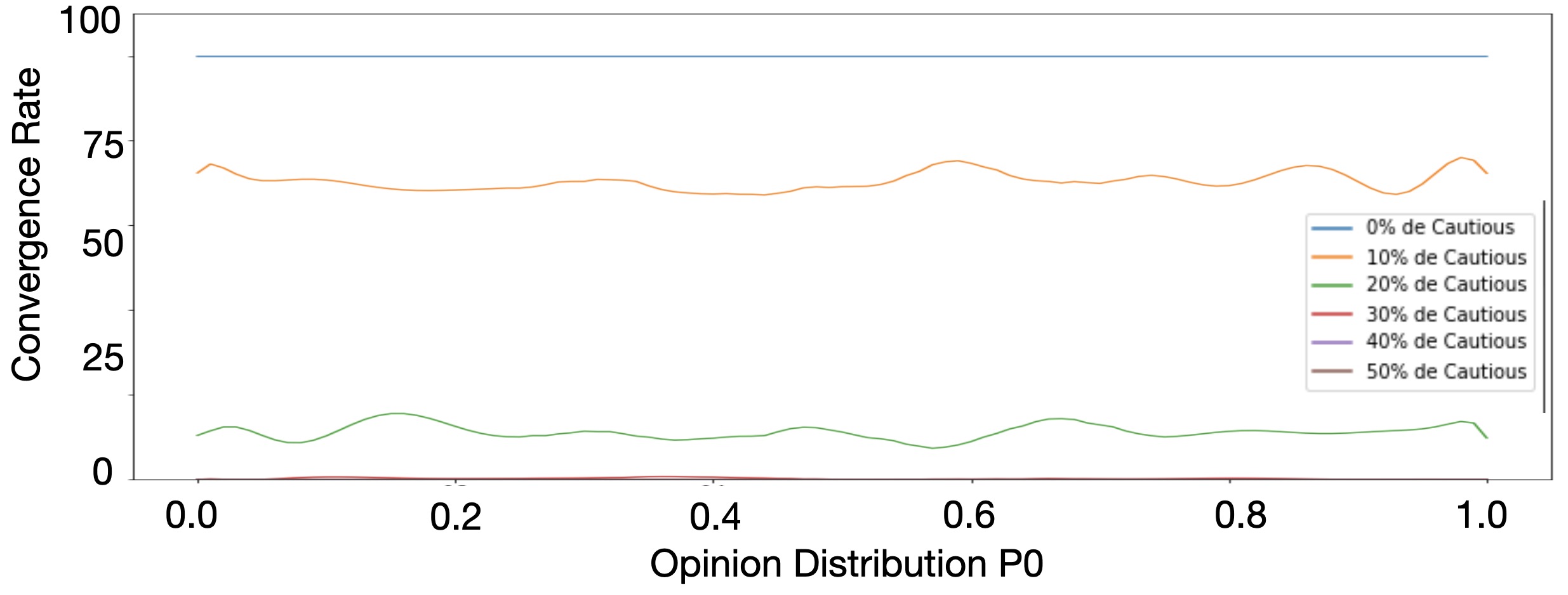} \\
    (c) Watts-Strogatz
    \caption{Convergence rate according to the initial division probability $P_0$ for different percentages of Cautious adversaries $P_{malicious}$}
    \label{P1_5}
\end{figure}

In Figure \ref{P1_5}, we vary the percentage of malicious nodes from $10\%$ up to $50\%$. We can clearly notice a huge difference between $P_{malicious} = 0\%$ and $P_{malicious} = 10\%$. When the network is corrupted with only 10\% of Cautious adversaries for both Grid and Torus,  the convergence rate drops  to 10\% and 20\%, respectively.  Watts-Strogatz resists up to 10 \% Cautious adversaries with a convergence rate of around 80\%. Even though Watts-Strogatz has a better resilience to Cautious than Grid and Torus, the convergence rate is  catastrophic for all topologies starting with  20\% of malicious nodes.


\subsubsection{Convergence rate according to the initial division probability $P_0$ for different network sizes, with 33\% Semi-Cautious Adversaries}

\begin{figure}[htbp]
    \centering
    \includegraphics[scale=0.09]{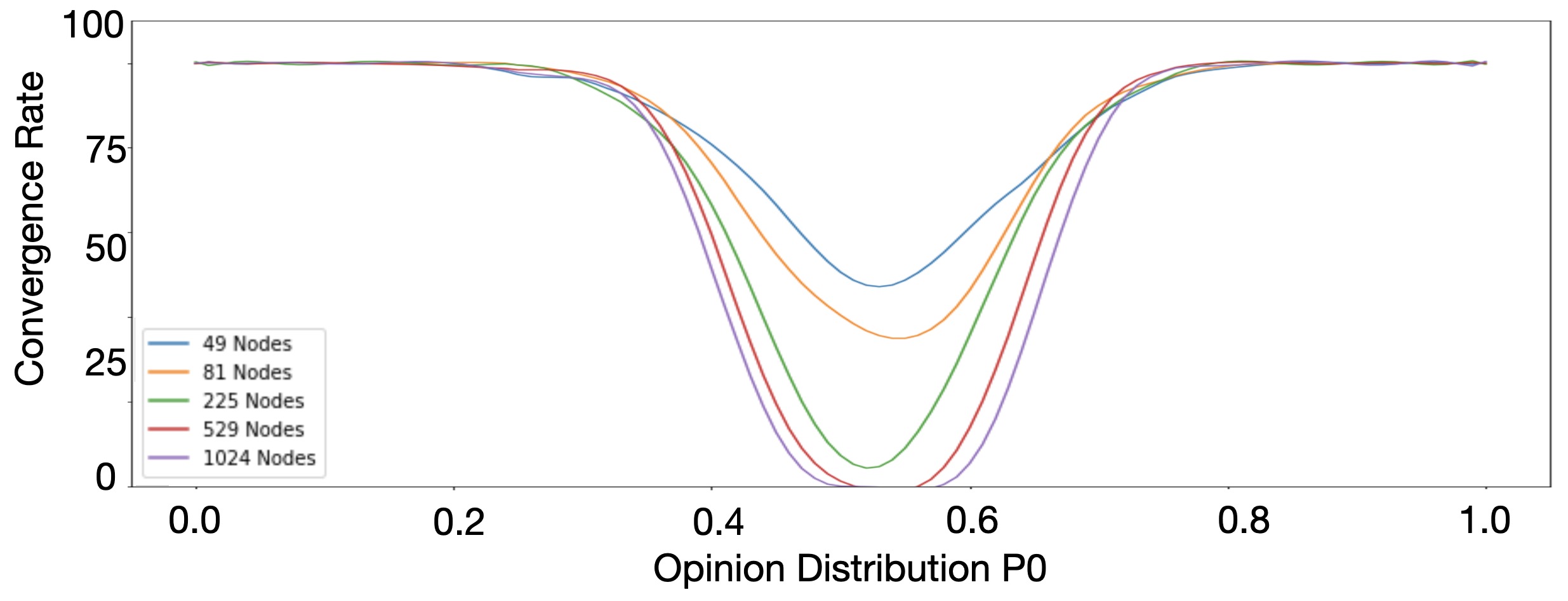}\\
    (a) 2D Grid\\
    \includegraphics[scale=0.09]{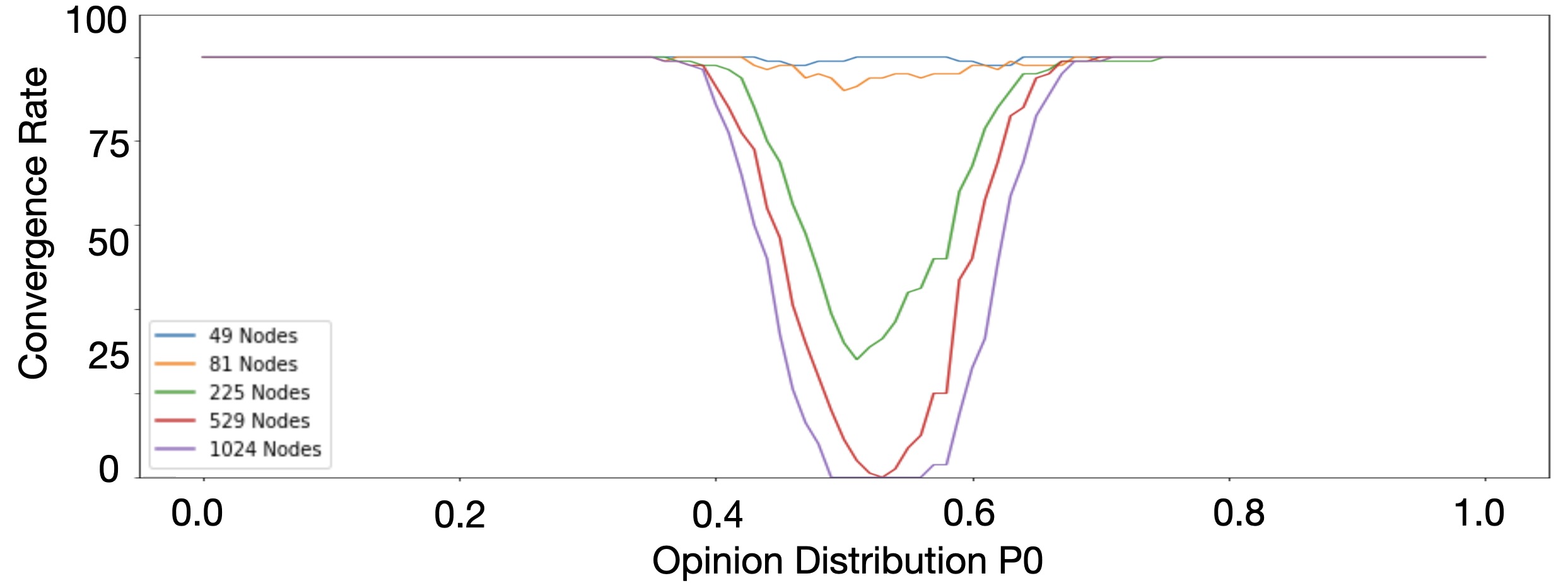} \\
    (b) Torus \\
    \includegraphics[scale=0.09]{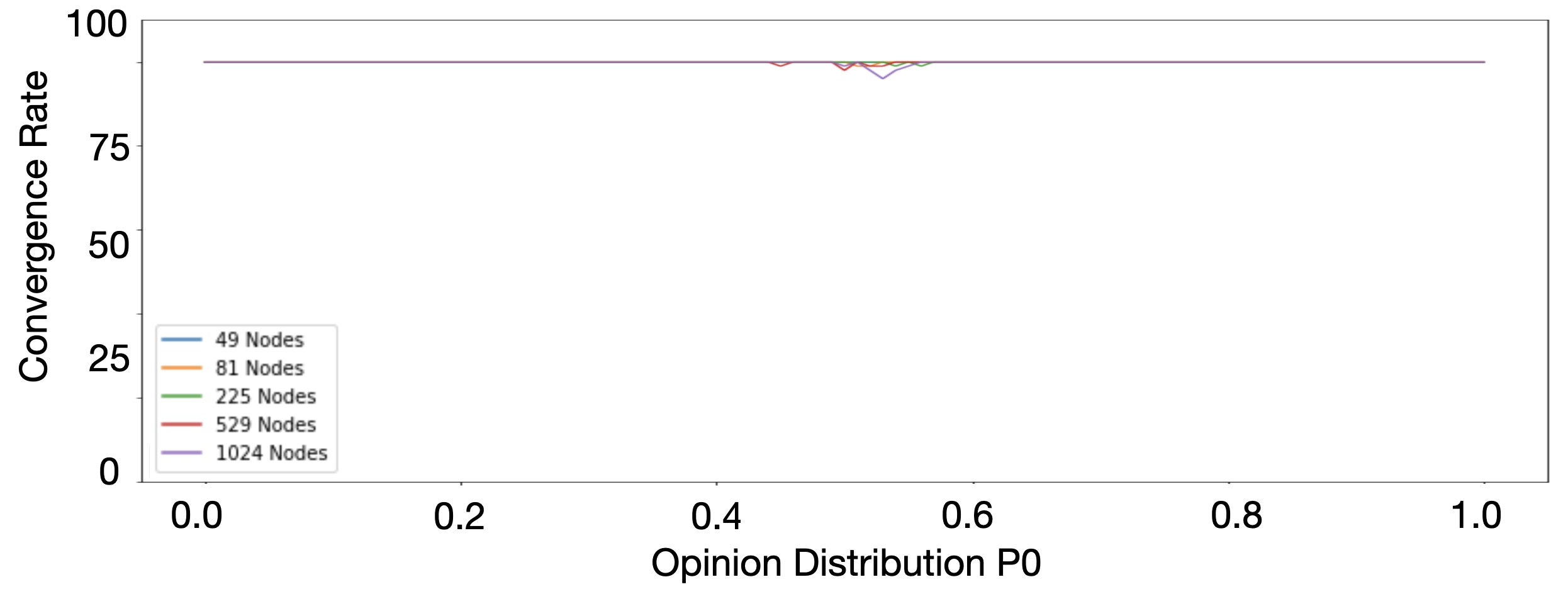} \\
    (c) Watts-Strogatz
    \caption{Convergence rate according to the initial division probability $P_0$ for different network sizes, with 33\% Semi-Cautious Adversaries}
    \label{P1_6}
\end{figure}


We an observe from Figure \ref{P1_6} that for each topology, Semi-Cautious adversaries do not have an important impact. It seems that the curves are the same as when there is no malicious nodes. This can be explained by the fact that Semi-Cautious adversaries do not lie, they just do not answer sometimes which does not corrupt the network. In theory, it should only slow down the network so the next graphs that we will see are going to make the percentage of Semi-Cautious vary and see if a big amount makes a difference or not.

\subsubsection{Convergence rate according to the initial division probability $P_0$ for different percentages of Semi-Cautious adversaries $P_{malicious}$}

\begin{figure}[htbp]
    \centering
    \includegraphics[scale=0.09]{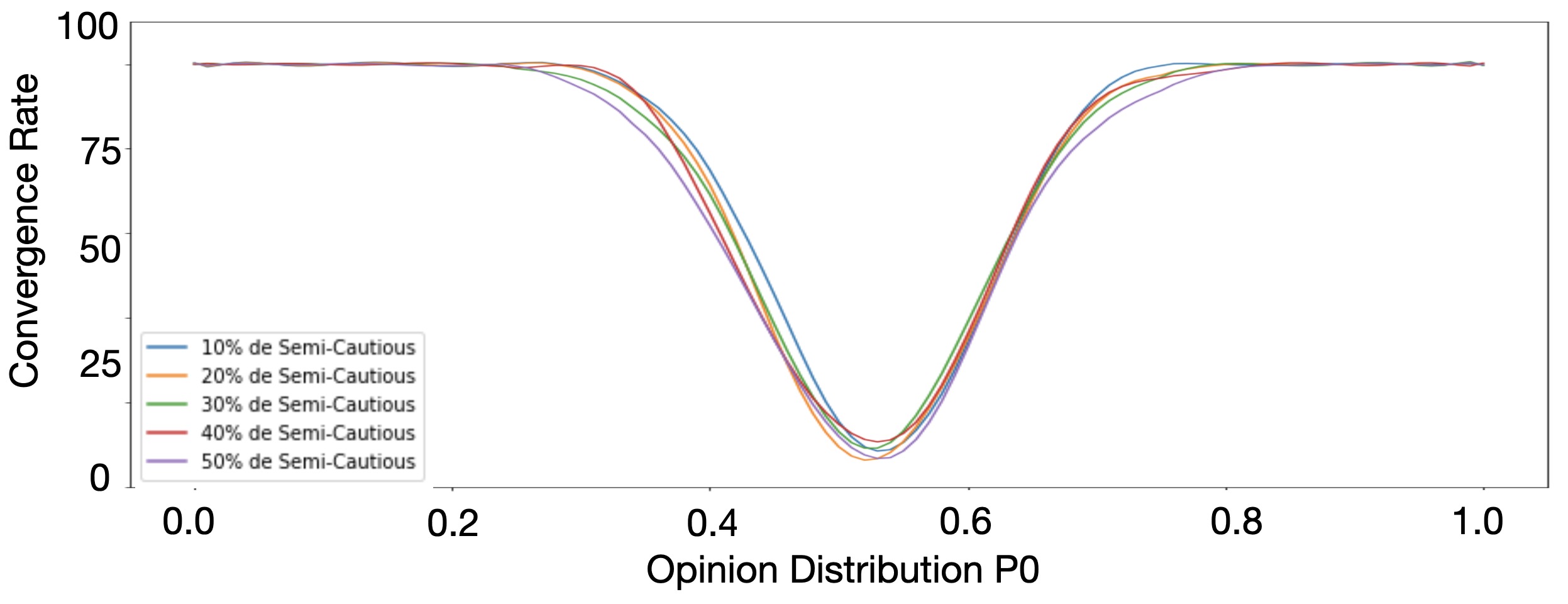}\\
    (a) 2D Grid\\
    \includegraphics[scale=0.09]{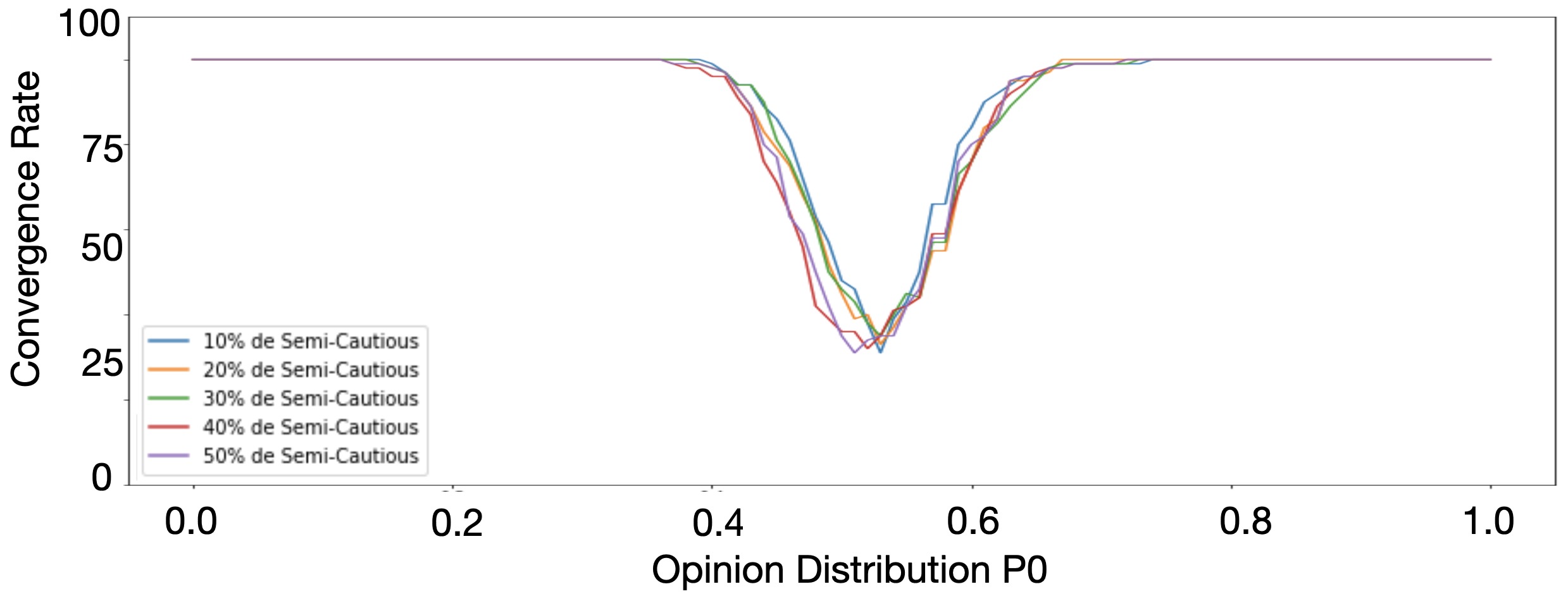} \\
    (b) Torus \\
    \includegraphics[scale=0.09]{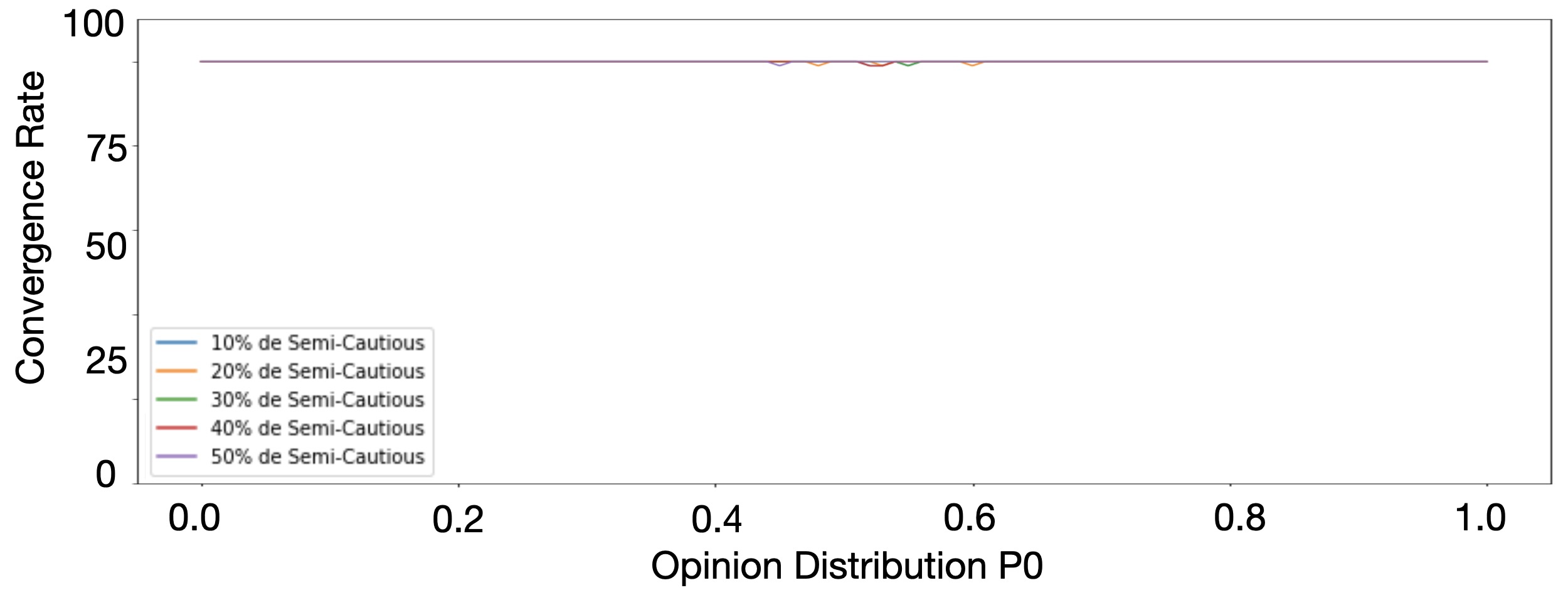} \\
    (c) Watts-Strogatz
    \caption{Convergence rate according to the initial division probability $P_0$ for different percentages of Semi-Cautious adversaries $P_{malicious}$}
    \label{P1_7}
\end{figure}

When the percentage of Semi-Cautious adversaries increases from $10\%$ to $50\%$, we do not see any notable difference in terms of convergence rate in Figure \ref{P1_7}, apart from a slight decrease. This means that the number of Semi-Cautious adversaries in the network does not affect the convergence, hence we can conclude on the fact that these types of malicious nodes are the least dangerous for the network. It also shows that if some devices are slow and they can't respond to the queries in time (this type of behavior is similar to being a Semi-Cautious adversary), the network will not be put at disadvantage because of these devices. 



\subsection{Cellular Consensus results}
Cellular Consensus (CC) has been evaluated using the same base configuration as Fast Probabilistic Consensus (FPC):
\begin{itemize}
    \item $K = 10$, $P = 1$ for Watts-Strogatz model;
    \item $P_{lying} = 50\%$ for Cautious;
    \item $P_{silence} = 50\%$ for Semi-Cautious;
    \item Number of rounds : $M = 30$;
\end{itemize}

\subsubsection{Convergence rate according to the initial division probability $P_0$ for different network sizes $N$, without malicious nodes}

\begin{figure}[htbp]
    \centering
    \includegraphics[scale=0.09]{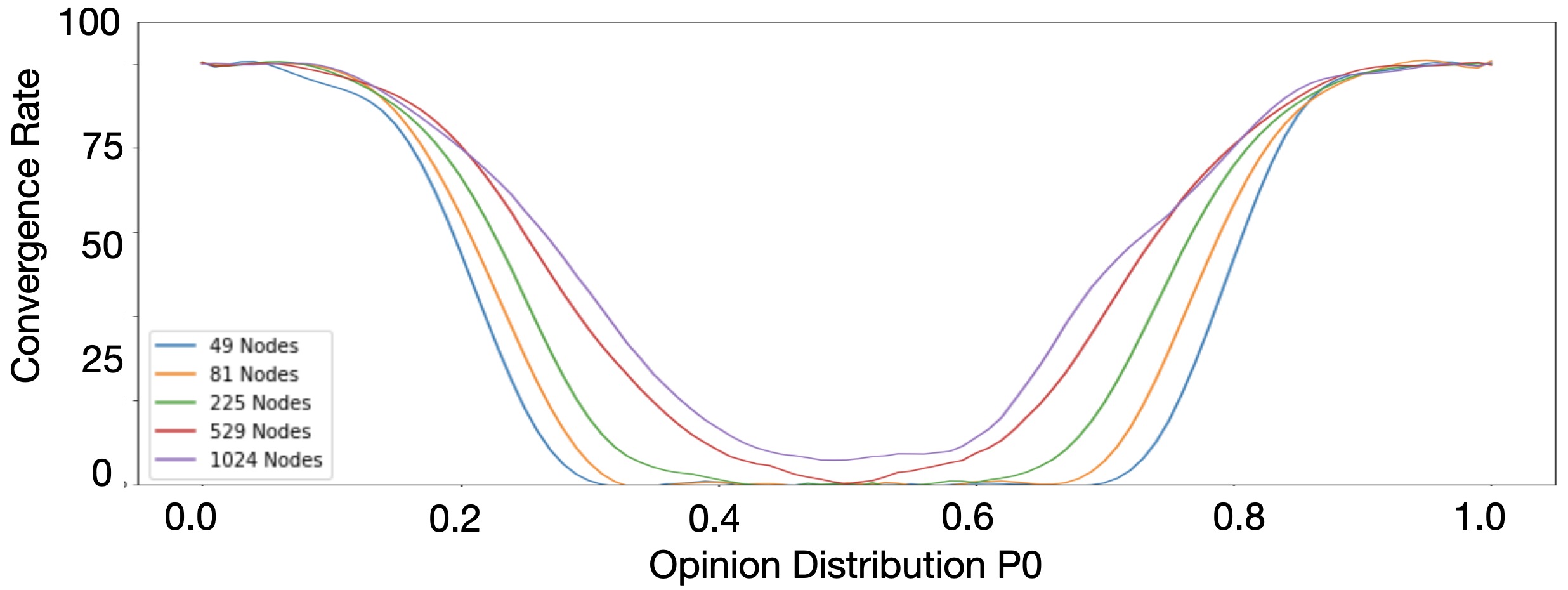}\\
    (a) 2D Grid\\
    \includegraphics[scale=0.09]{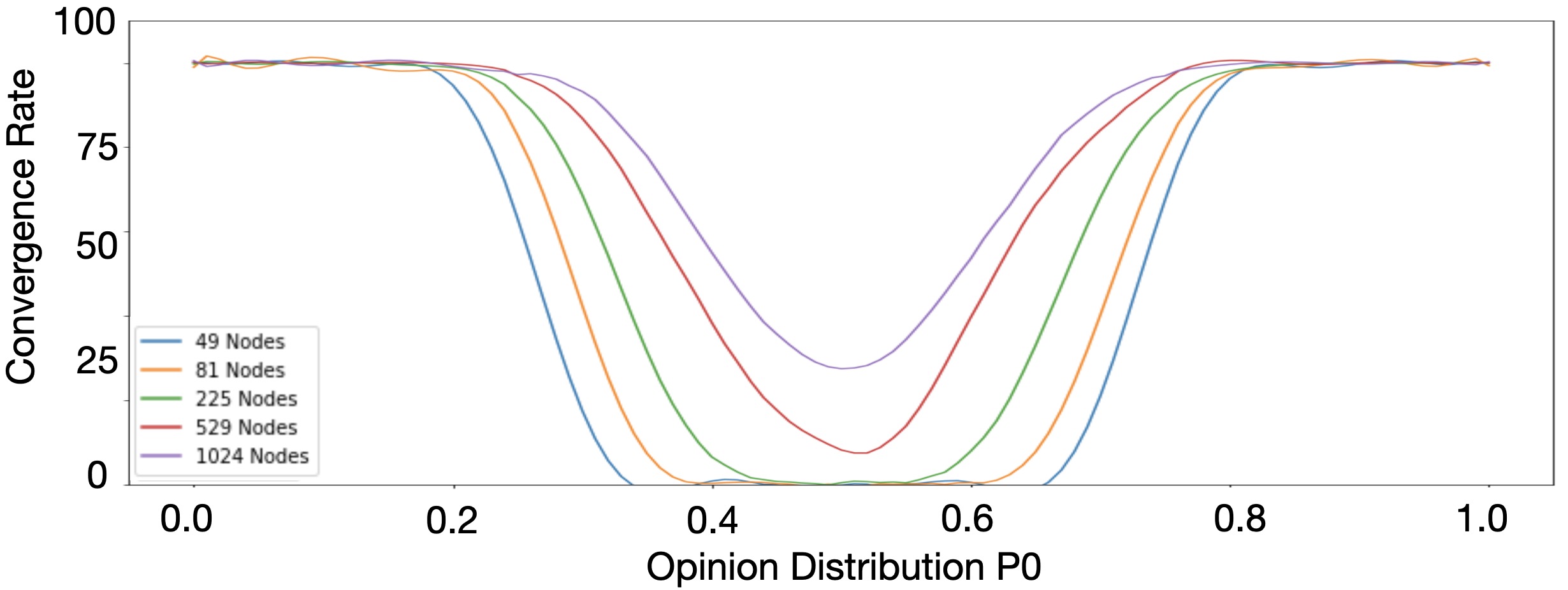} \\
    (b) Torus \\
    \includegraphics[scale=0.09]{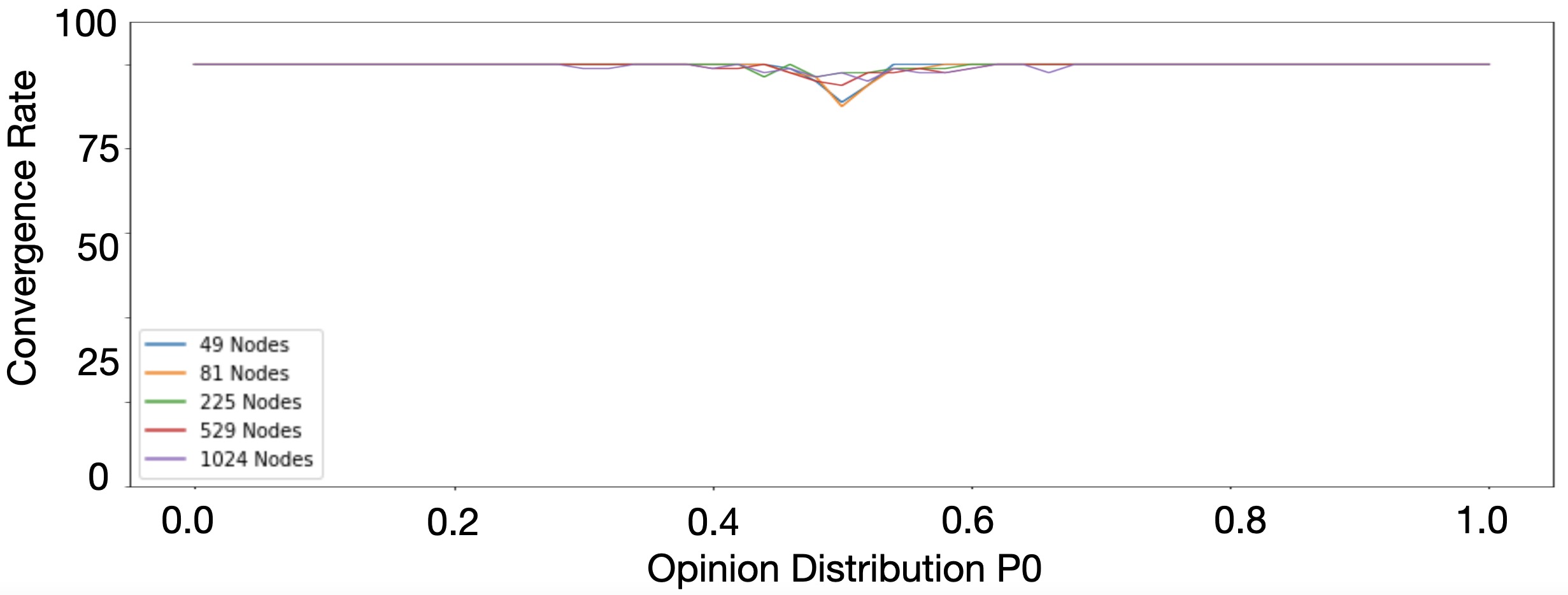} \\
    (c) Watts-Strogatz
    \caption{Convergence rate according to the initial division probability $P_0$ for different network sizes $N$, without malicious nodes}
    \label{P2_1}
\end{figure}

Through Figure \ref{P2_1}, for the three topologies, we observe that the number of nodes, $N$, is still an important parameter for the convergence of the network as in the case of Fast Probabilistic Consensus. Indeed, when the number of nodes increases, the convergence rate decreases for both Grid and Torus topologies. However, in terms of initial option distribution $P_0$, Cellular Consensus on Grid and Torus become more sensitive compared with  Fast Probabilistic Consensus. Convergence rates decrease drastically when $P_0 \in [0.25, 0.85]$ and never pass 50\%. In Watts-Strogatz model, Cellular Consensus has similar behavior as Fast Probabilistic Consensus and has a good scalability.

\subsubsection{Convergence rate according to the initial division probability $P_0$ for different average number of neighbors $K$, in Watts-Strogatz graph without malicious nodes}

\begin{figure}[htbp]
    \centering
    \includegraphics[scale=0.09]{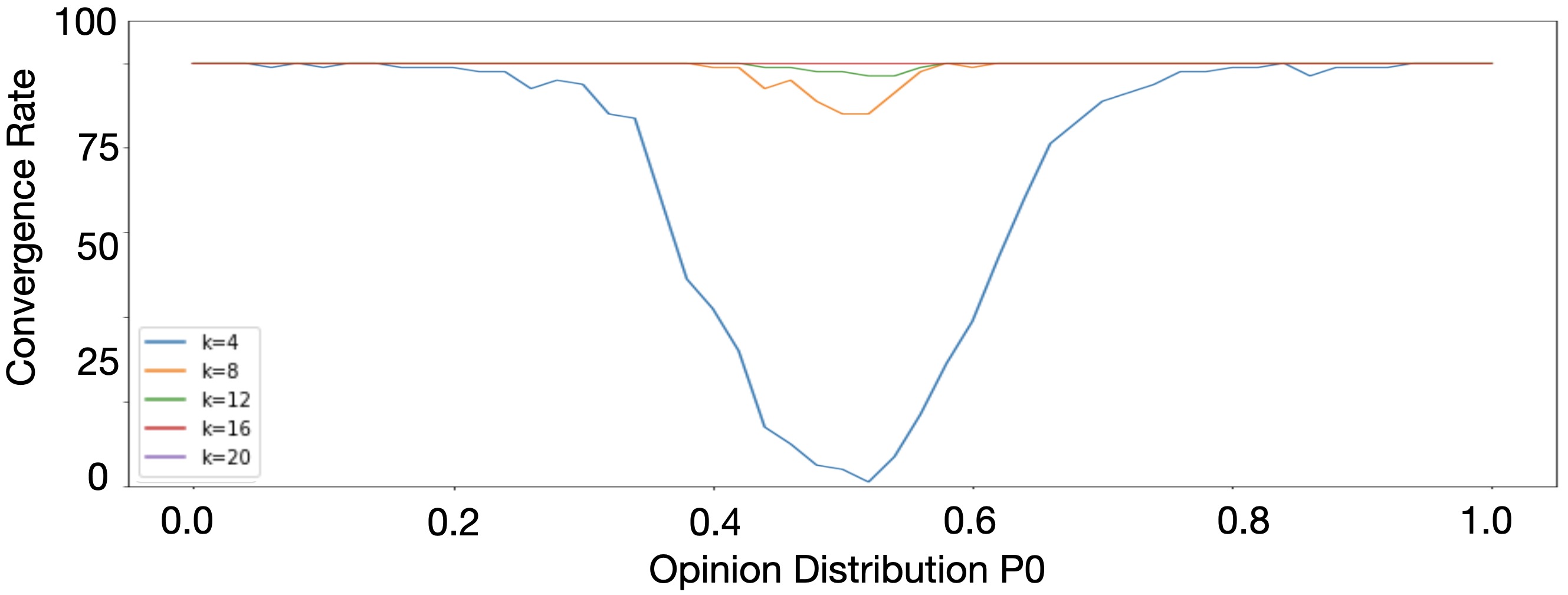}
    \caption{Convergence rate according to the initial division probability $P_0$ for different average number of neighbors $K$, in Watts-Strogatz graph without malicious nodes}
    \label{P2_2}
\end{figure}

Contrary to Fast Probabilistic Consensus, in Figure \ref{P2_2}, we show that adjusting $K$ for Watts-Strogatz topology impacts clearly the convergence rate in Cellular Consensus: when $K$ gets smaller, Watts-Strogatz topology becomes more sensitive to the initial opinion distribution $P_0$ and networks size $N$. We therefore test different $K$ values for the Watts-Strogatz model  to evaluate the resilience of Cellular Consensus  to Cautious and Semi-Cautious Adversaries. Recall that Cellular Consensus is based on the communication between neighbors and that we have the average number of neighbors $K < 4$ for Grid, $K = 4$ for Torus and $K = 10$ for Watts-Strogatz in our simulations. We conclude here that without malicious nodes, the higher the average number of neighbors is, the better the convergence rate will be. 



\subsubsection{Convergence rate according to the initial division probability $P_0$ for different network sizes $N$, with 33\% Cautious Adversaries}

\begin{figure}[htbp]
    \centering
    \includegraphics[scale=0.09]{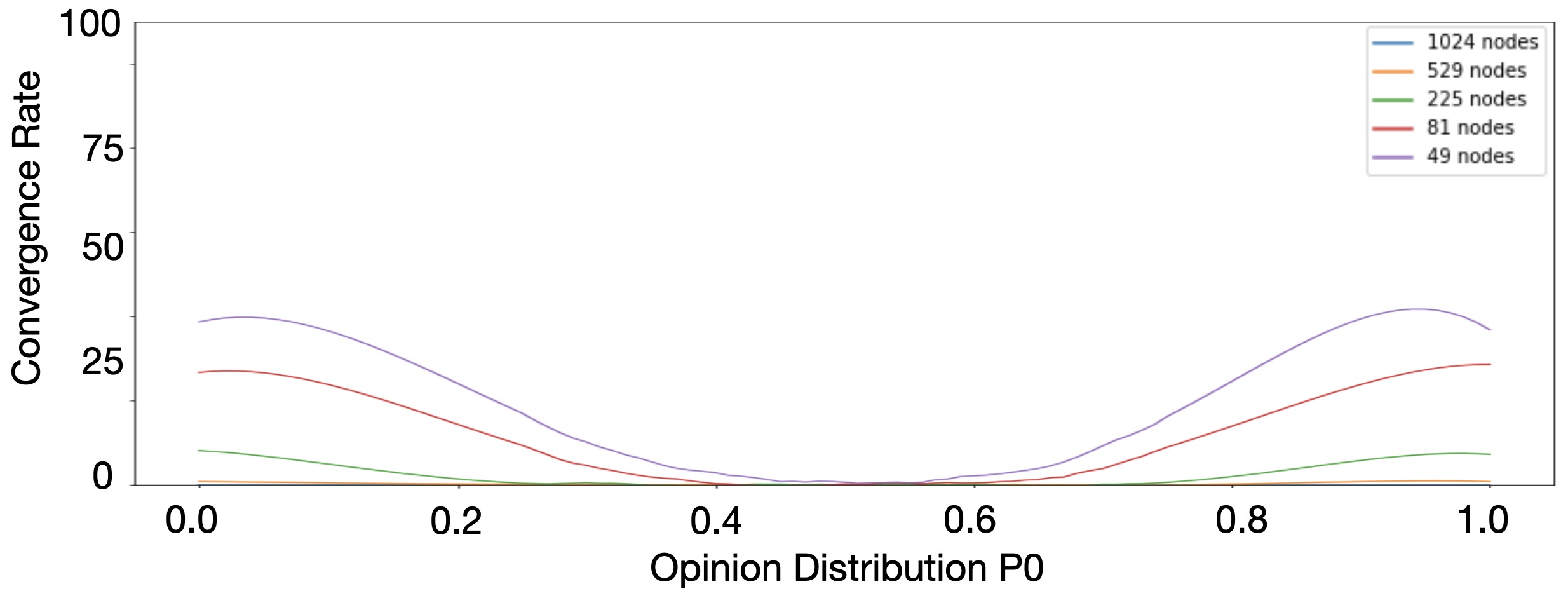}\\
    (a) 2D Grid\\
    \includegraphics[scale=0.09]{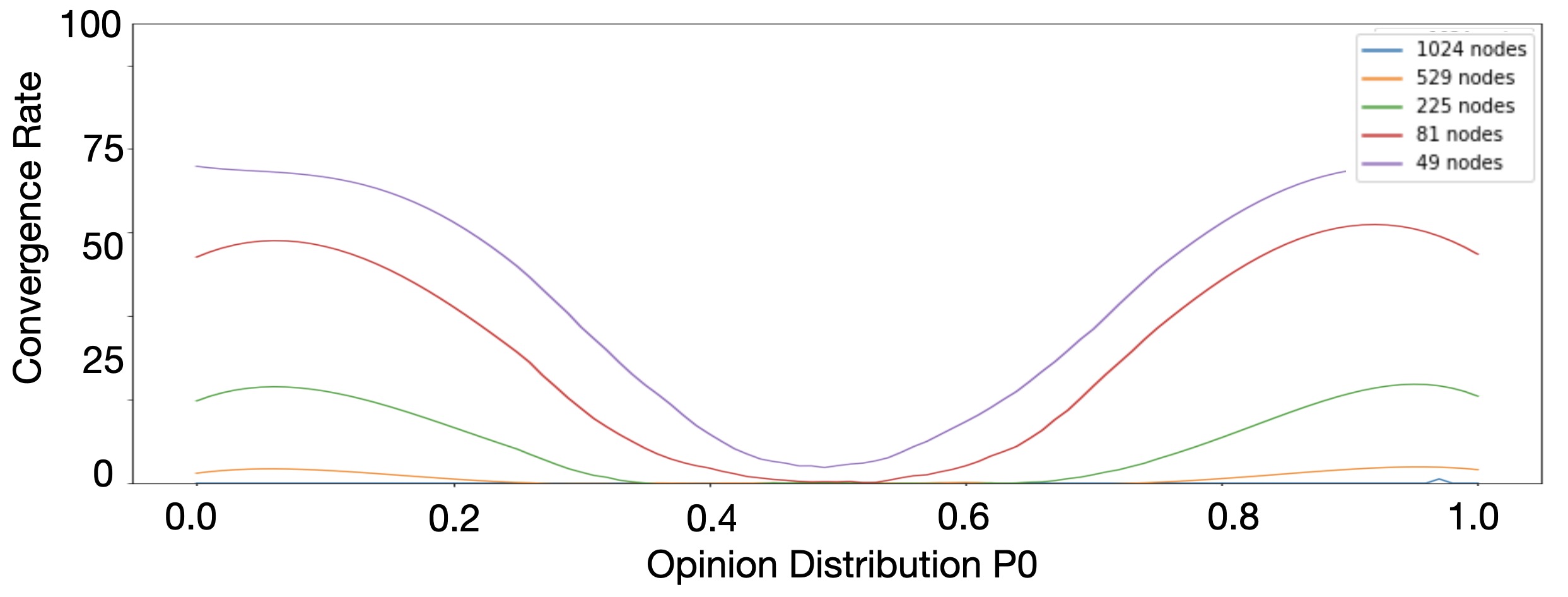} \\
    (b) Torus \\
    \includegraphics[scale=0.09]{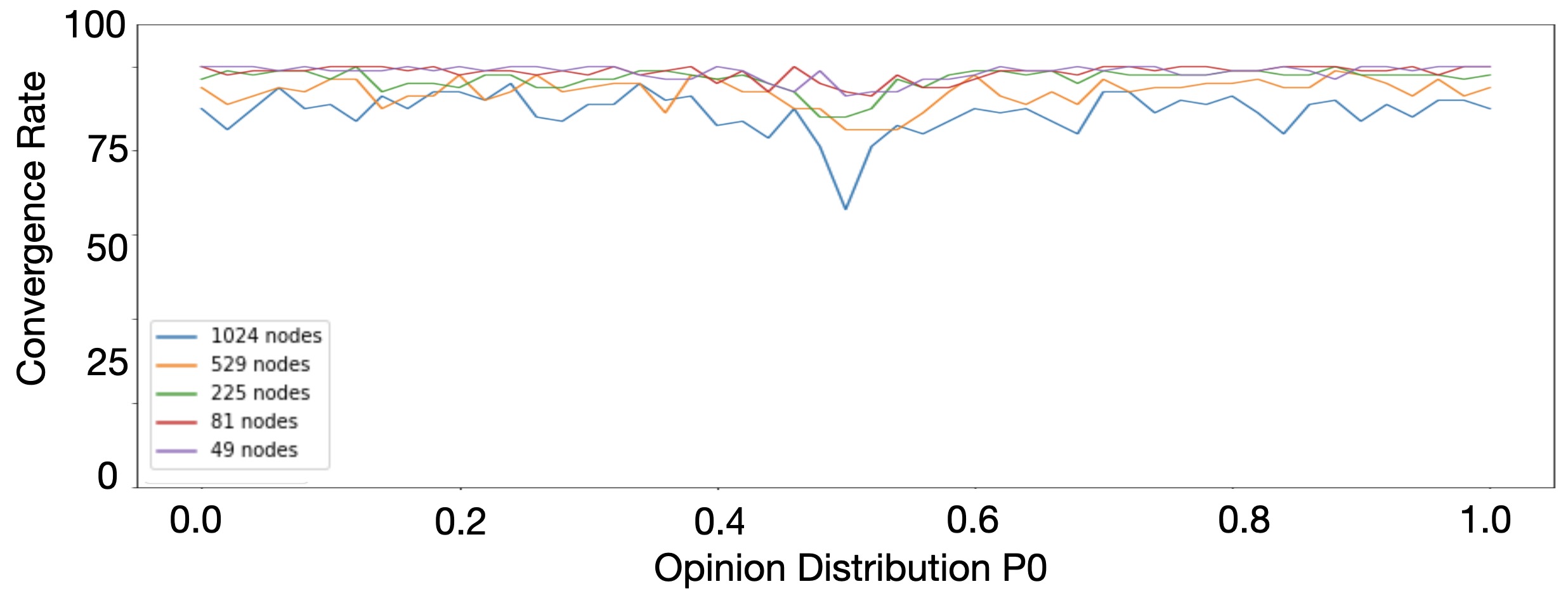} \\
    (c) Watts-Strogatz
    \caption{Convergence rate according to the initial division probability $P_0$ for different network sizes $N$, with 33\% Cautious Adversaries}
    \label{P2_3}
\end{figure}

First we study the convergence rate when $N/3$ (33\%) nodes are Cautious adversaries.
When we introduce Cautious adversaries in Cellular Consensus, we observe, in Figure \ref{P2_3}, the same tendency as without malicious nodes by adjusting network size $N$ however, the convergence rate  is much lower. Indeed when the malicious nodes are detected, they are immediately blacklisted which means that some nodes lose neighbors, which leads to this decrease. However, compared with Fast Probabilistic Consensus, Cellular Consensus has a better resilience to Cautious Adversaries since it can detect malicious nodes and block them once detected. 
In the sequel,  we fix the network size $N = 225$  as in Fast Probabilistic Consensus. 

\subsubsection{Convergence rate according to the initial division probability $P_0$ for different average number of neighbors $K$, in Watts-Strogatz graph with 33\% Cautious adversaries}

\begin{figure}[htbp]
    \centering
    \includegraphics[scale=0.09]{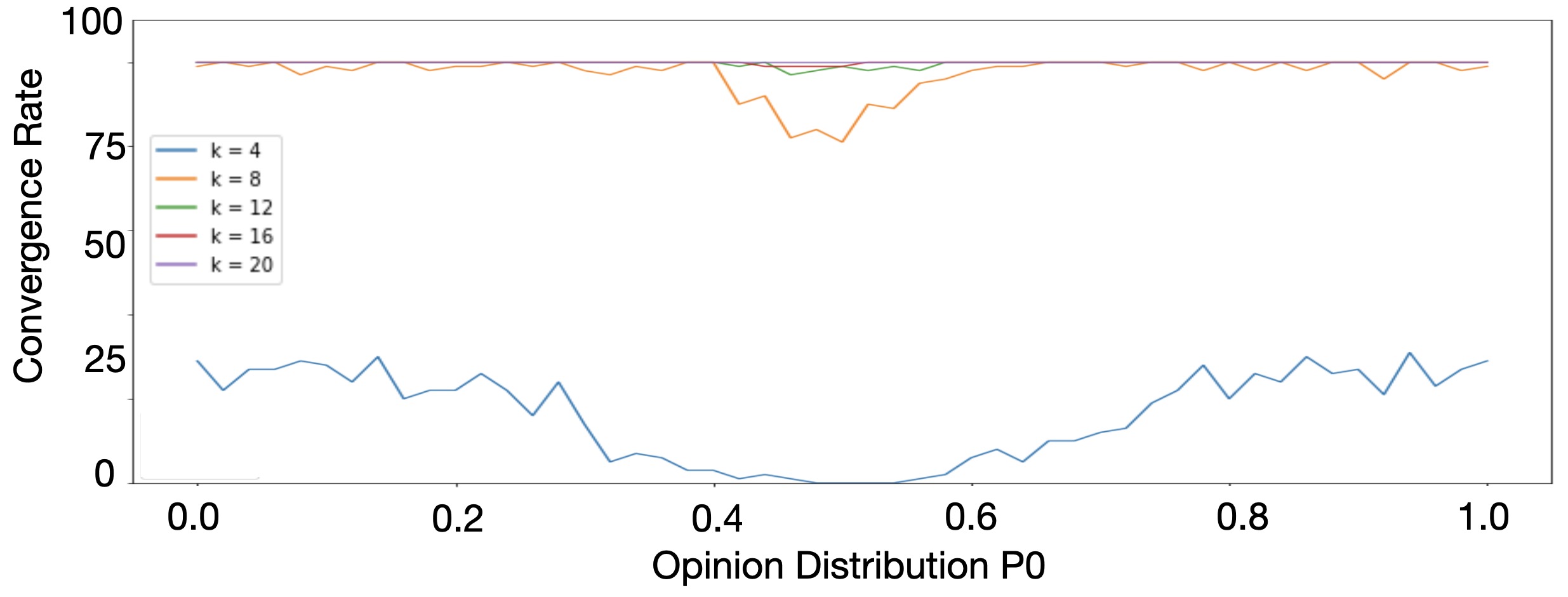}
    \caption{Convergence rate according to the initial division probability $P_0$ for different average number of neighbors $K$, in Watts-Strogatz graph with Cautious adversaries}
    \label{P2_4}
\end{figure}

In the following we study the convergence rate of Cellular Consensus in Watts-Strogatz  model when varying $K$ and considering N/3 (33\%) Cautious adversaries. In Figure \ref{P2_4}, the results confirm that the average number of neighbors influences the convergence rate. 

\subsubsection{Convergence rate according to the initial division probability $P_0$ for different percentages of Cautious adversaries $P_{malicious}$}

\begin{figure}[htbp]
    \centering
    \includegraphics[scale=0.09]{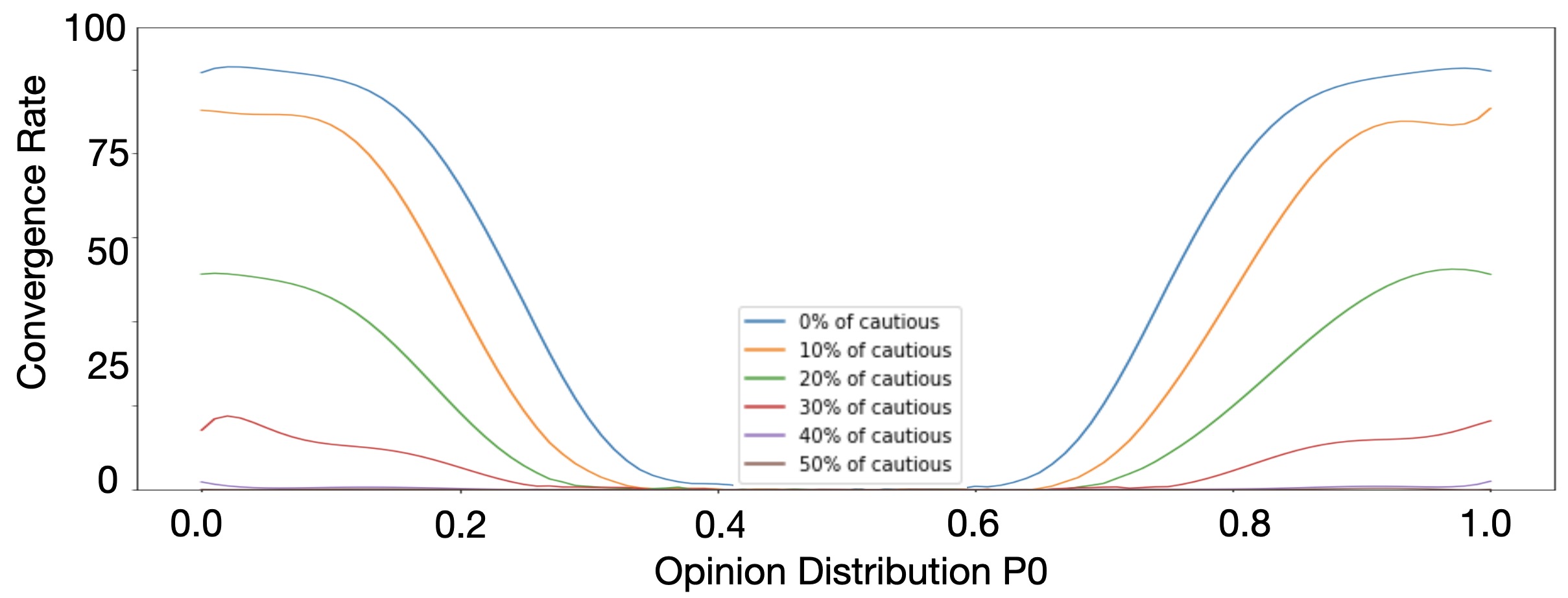}\\
    (a) 2D Grid\\
    \includegraphics[scale=0.09]{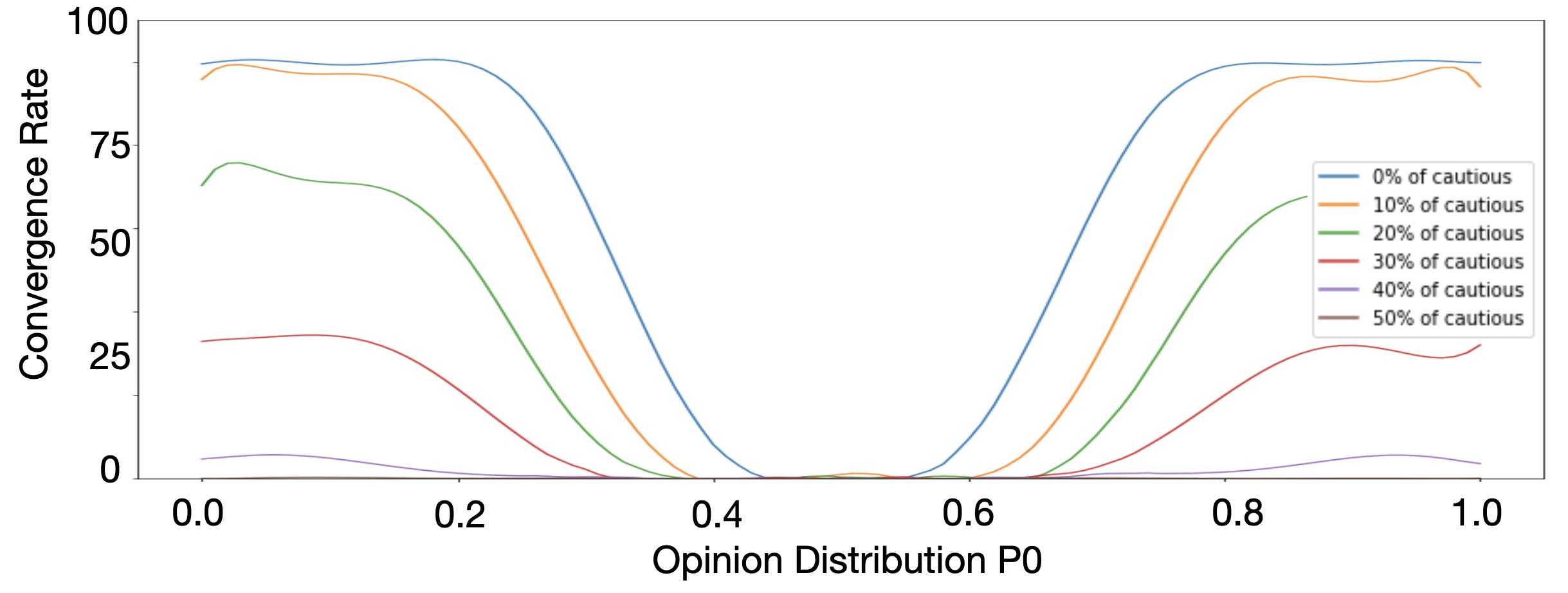} \\
    (b) Torus \\
    \includegraphics[scale=0.09]{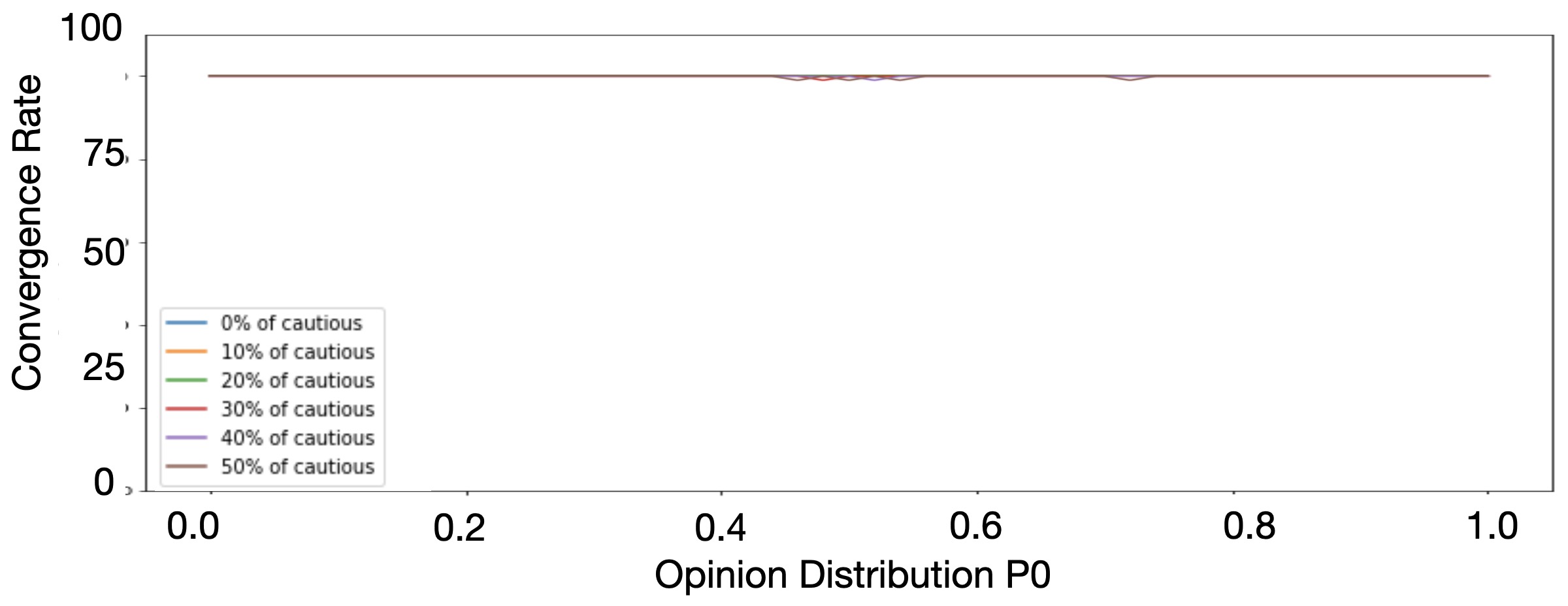} \\
    (c) Watts-Strogatz
    \caption{Convergence rate ccording to the initial division probability $P_0$ for different percentages of Cautious adversaries $P_{malicious}$}
    \label{P2_5}
\end{figure}

When  $P_{malicious}$ varies, we observed in \ref{P2_5} a decrease in the convergence rate. We observe that the results obtained with Torus are better than those in Grid.  As for the results obtained with the Watts-Strogatz topology, they are consistently good and sometimes even excellent with $K = 20$. 

\subsubsection{Convergence rate according to the initial division probability $P_0$ for different network sizes $N$, with 33\% Semi-Cautious Adversaries}

\begin{figure}[htbp]
    \centering
    \includegraphics[scale=0.09]{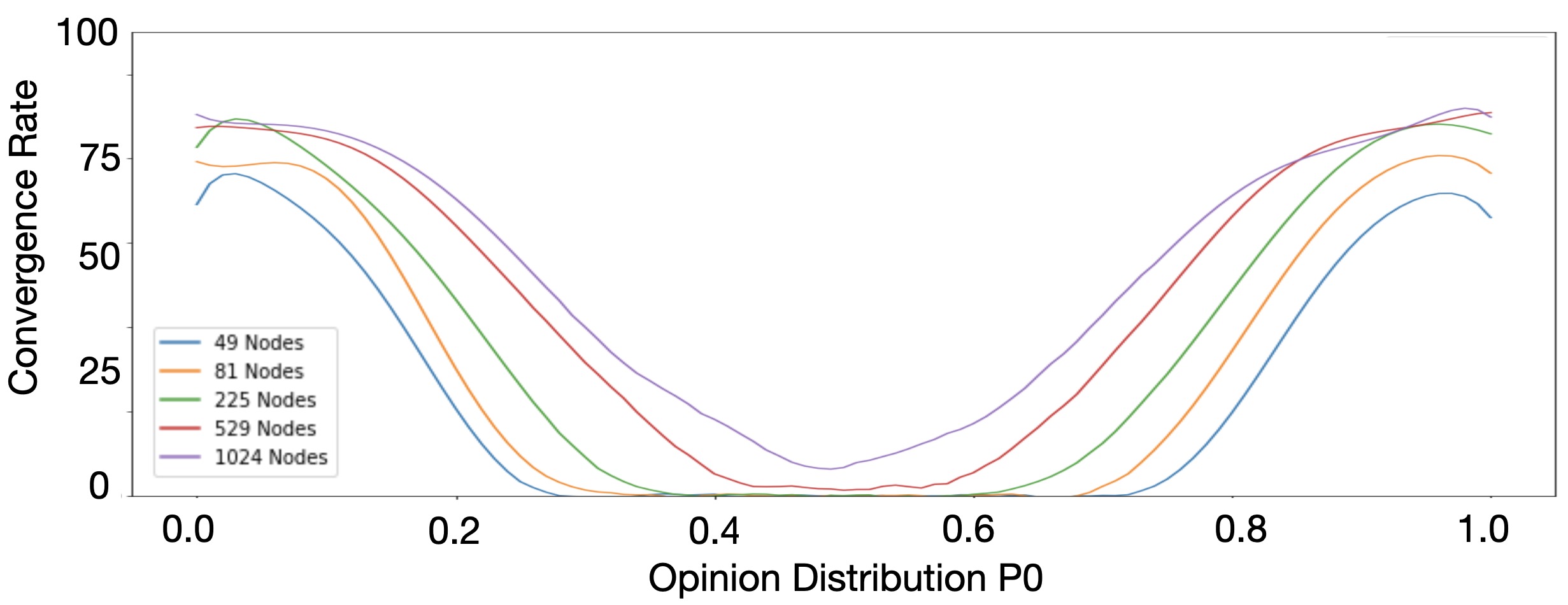}\\
    (a) 2D Grid\\
    \includegraphics[scale=0.09]{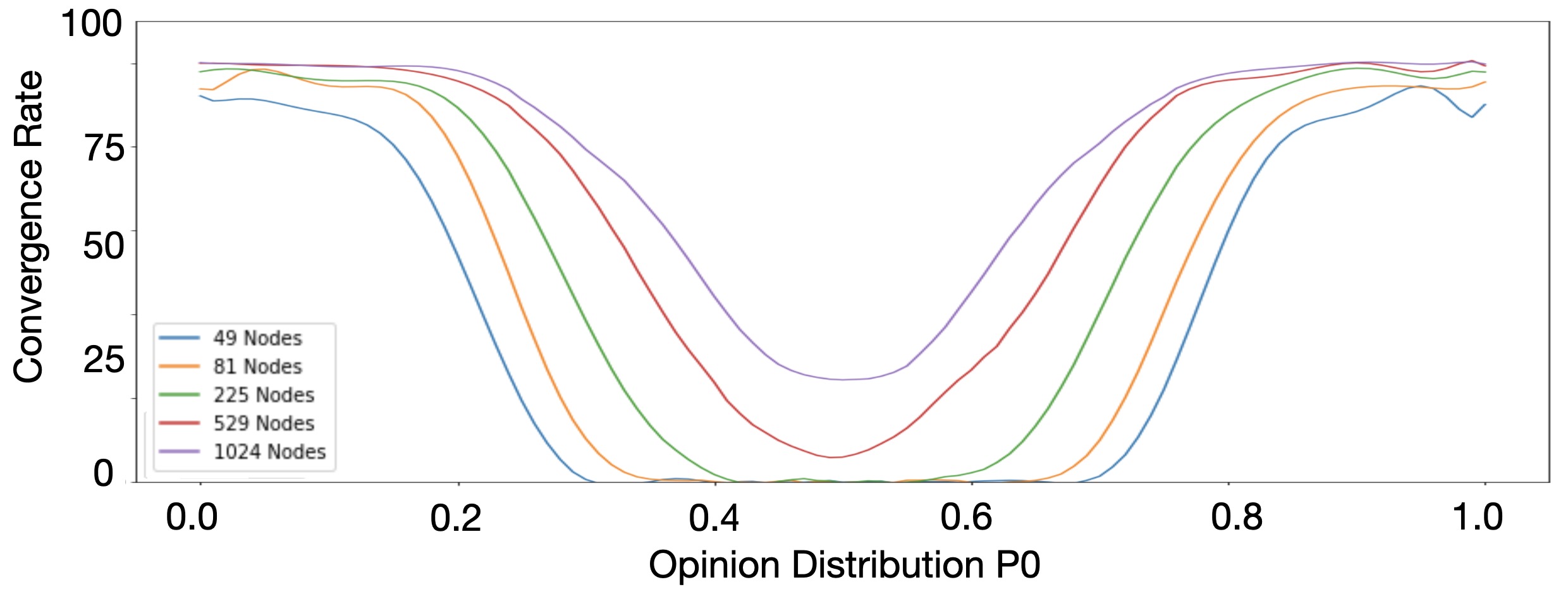} \\
    (b) Torus \\
    \includegraphics[scale=0.09]{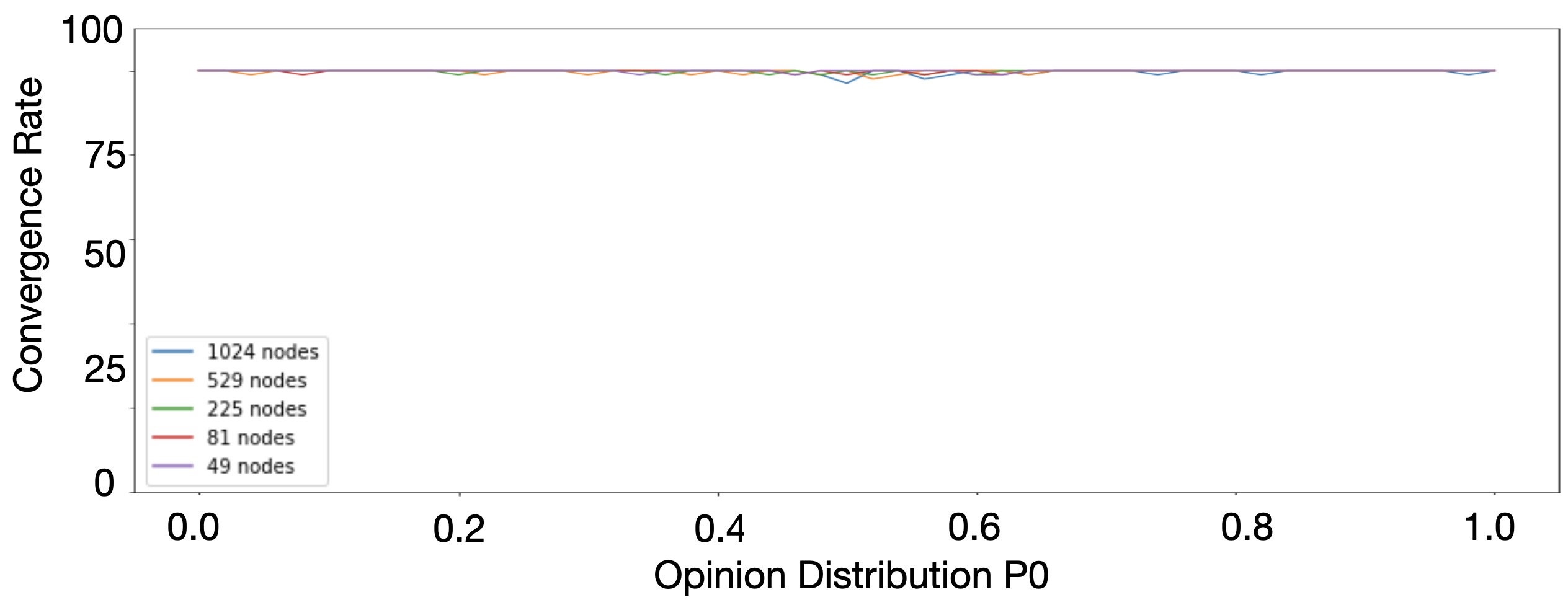} \\
    (c) Watts-Strogatz
    \caption{Convergence rate according to the initial division probability $P_0$ for different network sizes $N$, with 33\% Semi-Cautious Adversaries}
    \label{P2_6}
\end{figure}

When we introduce Semi-Cautious adversaries in the network, we observe in Figure \ref{P2_6} that when $0,3 \leq P_0 \leq 0,7$, the convergence rate is the same than the case without malicious nodes for 2D Grid and Torus topologies. In addition, for Watts-Strogatz topology, when $P_0$ is in this interval, we have better results that without byzantines nodes. This phenomenon is explained by the fact that when a network is in disagreement, the nodes which do not respond, make it possible to take a decision on the opinion, and therefore to promote convergence. It seems that this process, appears only when the average number of neighbors is high. However, when $0 \leq P_0 < 0,3$ or $0,7 < P_0 \leq 1$, the convergence rate is lower than normal, which seems quite logical. Indeed, when the network is in agreement, the Semi-Cautious adversaries disturb him and cause a drop of the convergence rate. 

\subsubsection{Convergence rate according to the initial division probability $P_0$ for different average number of neighbors $K$, in Watts-Strogatz graph with 33\% Semi-Cautious adversaries}

\begin{figure}[htbp]
    \centering
    \includegraphics[scale=0.09]{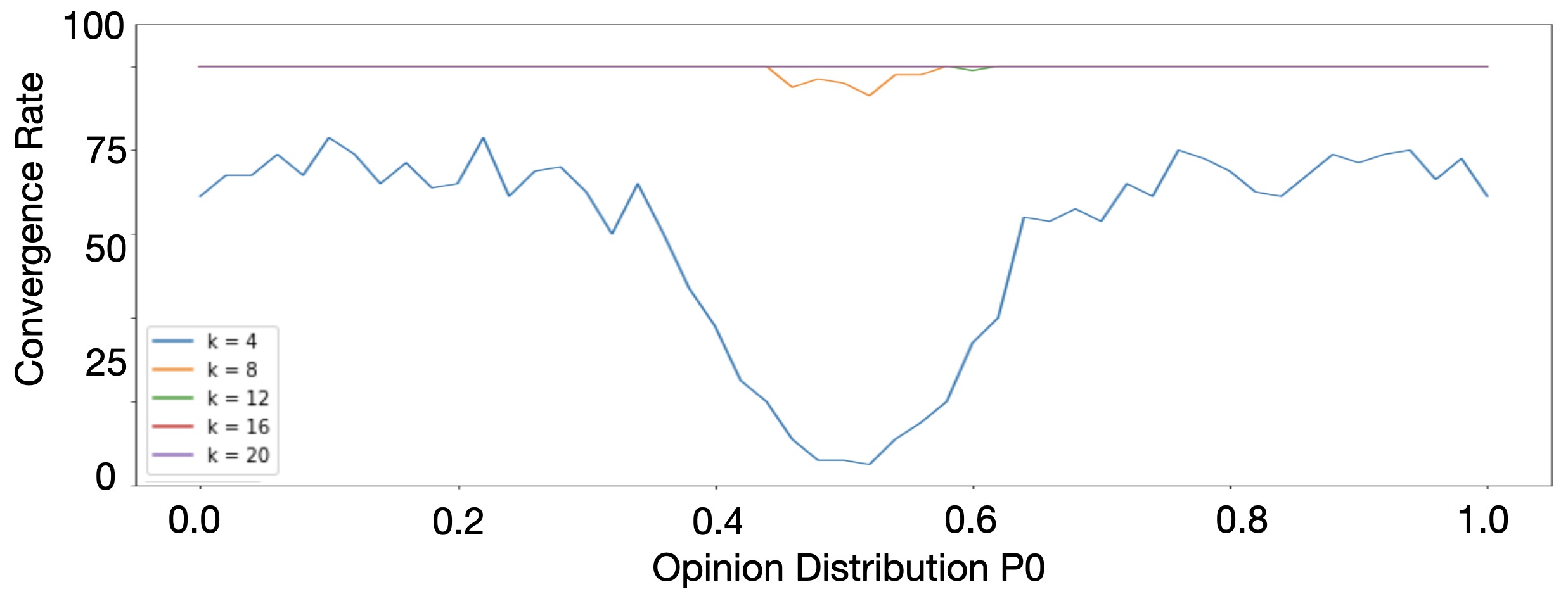}
    \caption{Convergence rate according to the initial division probability $P_0$ for different average number of neighbors $K$, in Watts-Strogatz graph with Semi-Cautious adversaries}
    \label{P2_7}
\end{figure}

The results in Figure \ref{P2_7}  show that when the average number of neighbors is low, $K <= 4$, the convergence rate is relatively low for $0 \leq P_0 < 0,3$ or $0,7 < P_0 \leq 1$ and similar to a network without malicious nodes for $0,3 \leq P_0 \leq 0,7$. When the average number of neighbors is higher, the results can be better. Same as in Cautious case, by adjusting $K$, Watts-Strogatz might have a good resilience to this type of attack.


\subsubsection{Convergence rate according to the initial division probability $P_0$ for different percentages of Semi-Cautious adversaries $P_{malicious}$}

\begin{figure}[htbp]
    \centering
    \includegraphics[scale=0.09]{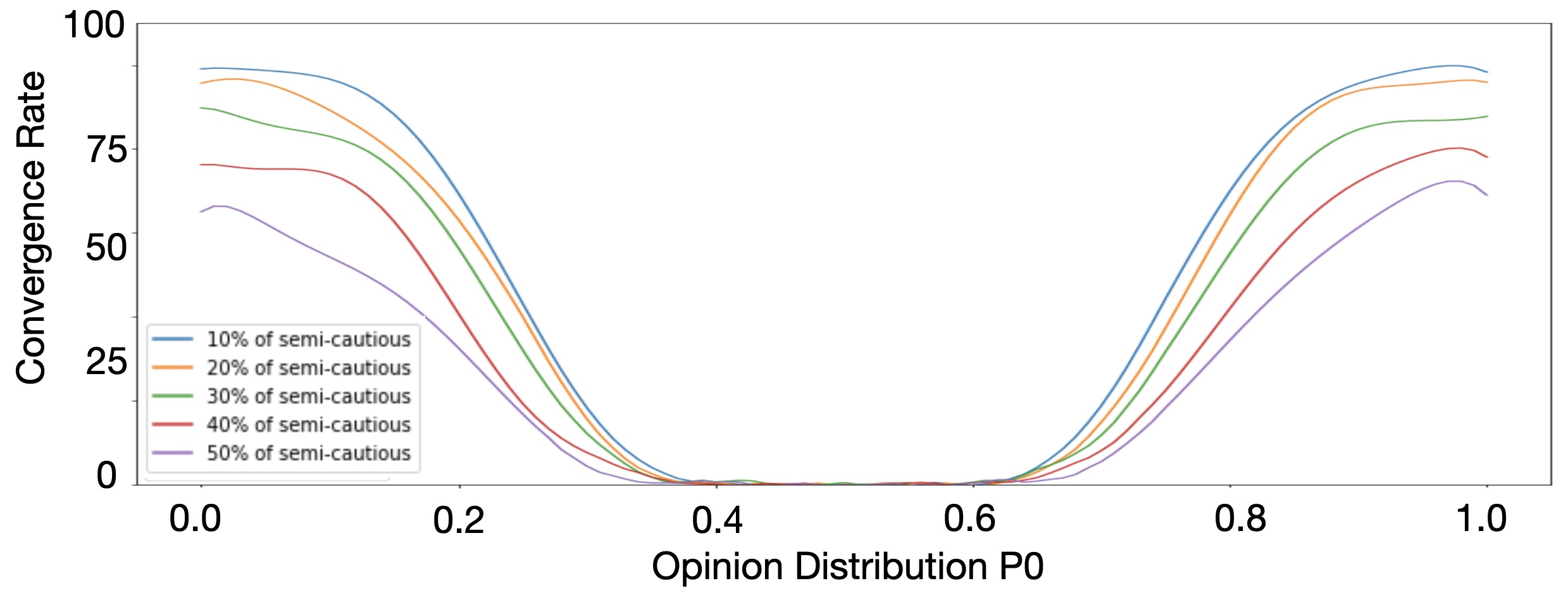}\\
    (a) 2D Grid\\
    \includegraphics[scale=0.09]{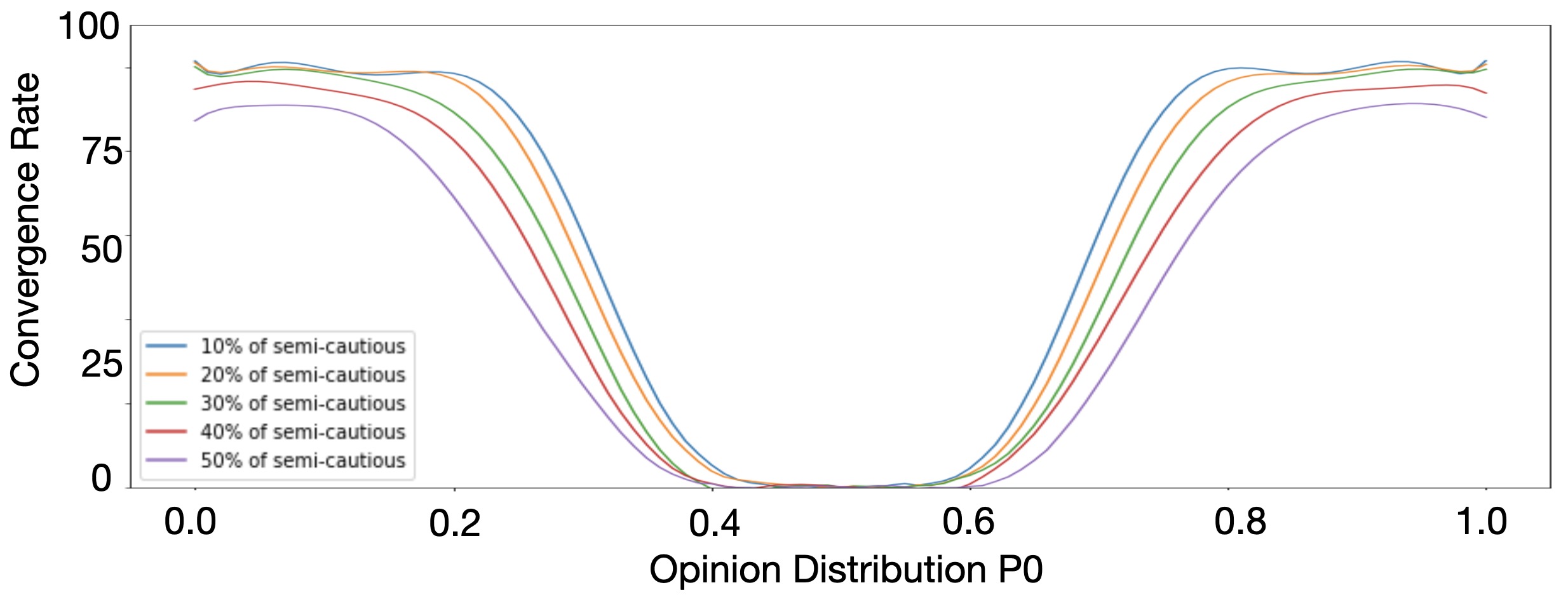} \\
    (b) Torus \\
    \includegraphics[scale=0.09]{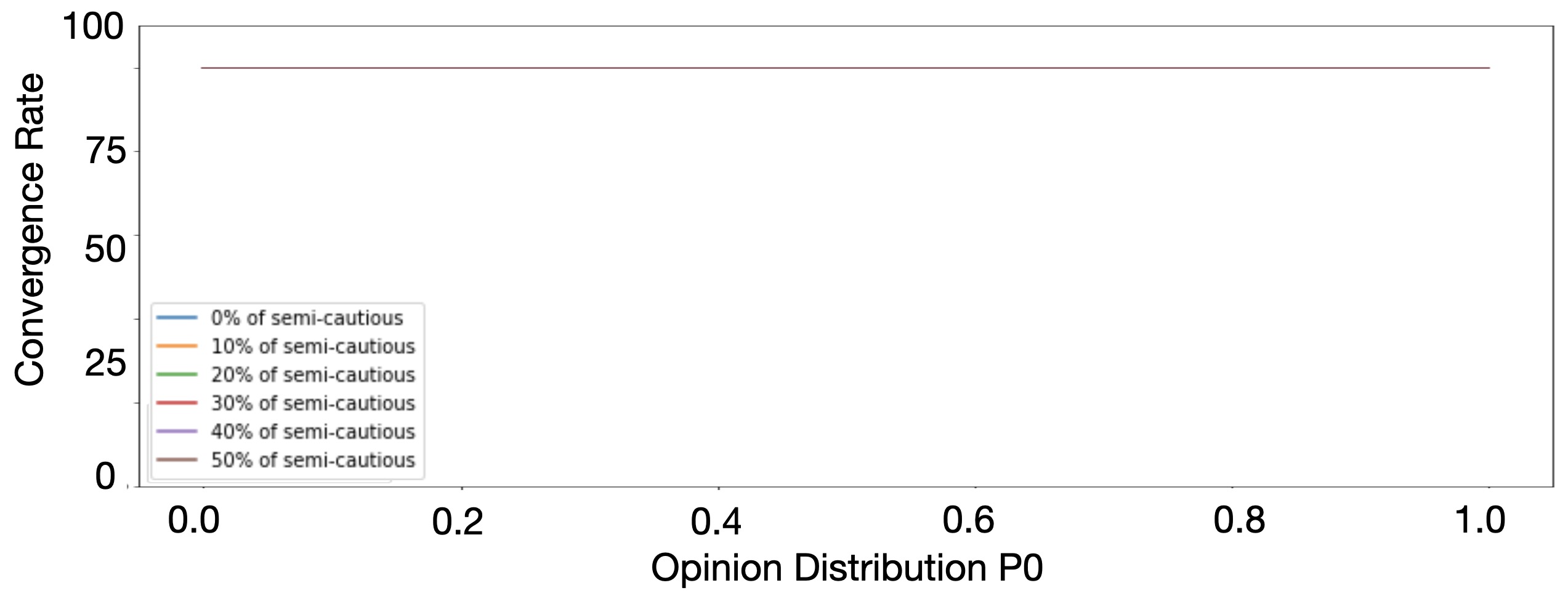} \\
    (c) Watts-Strogatz
    \caption{Convergence rate according to the initial division probability $P_0$ for different percentages of Semi-Cautious adversaries $P_{malicious}$}
    \label{P2_8}
\end{figure}


When we vary the $P_{malicious}$ of Semi-Cautious adversary in Figure \ref{P2_8}, we observe a decrease in convergence which seems quite logical. By comparing the different topologies, we observe that the results obtained with the Torus are better than those obtained with the 2D grid. As for the results obtained with the Watts-Strogatz topology, they are consistently excellent with $k = 20$. As we have seen in the case of Cautious adversaries, we see that when the number of byzantines nodes is low ($0\% \leq P_{malicious} \leq 10\%$) or high ($30\% \leq P_{malicious} \leq 50\%$) the convergence rate does not evolve a lot. 




\section{Conclusion}
In this paper we extensively evaluate the resilience of  Fast Probabilistic Consensus and Cellular Consensus (two agreement building blocks introduced for  IOTA blockchain \cite{coordicide}). Our evaluation focused on the impact of the underlying network topology on the convergence rate of the algorithms and their resilience to various adversaries (cautious, semi-cautious and Berserk).  
We showed that the initial opinion state distribution of each node and the total number of nodes in the network will seriously affect the convergence results, especially in 2D Grid and Torus topologies. These effects will not be alleviated by increasing the run time.

In Fast Probabilistic Consensus, Cautious and Berserk adversaries will cause serious  convergence issues even with low power adversaries. However, semi-cautious adversaries seem to not be a menace.  

In the Cellular Consensus, all three adversaries have an impact on the convergence, but none of them is as serious as in the case of Fast Probabilistic Consensus. 

Interestingly, the Watts-Strogatz topology when the density is properly adjusted can even eliminate the effects of adversaries.

{\color{red}We plan to continue this work by giving mathematical modeling and explaining our results formally.}

\bibliographystyle{IEEEtran}
\bibliography{sample-article,green-dlt,references_CNAM,sample-dmtcs}

\end{document}